\documentclass[aip,pof,floatfix,onecolumn,reprint]{revtex4-1}

\usepackage{graphicx}
\usepackage{amsmath}
\usepackage{amsfonts}
\usepackage{amssymb}
\usepackage{subfigure}
\usepackage{natbib}

\usepackage{color}

\newcommand{\strain}{\lambda}
\newcommand{\taylorscale}{\lambda_g}
\newcommand{\dissip}{\varepsilon}
\newcommand{\cross}{\times}

\newcommand{\ie}{\textit{i.e.}}

\begin{document}

\title{Tetrahedron deformation and alignment of perceived vorticity and strain in a turbulent flow }

\author{Alain Pumir}
\affiliation{Laboratoire de Physique, Ecole Normale Sup\'erieure de Lyon, Universit\'e Lyon 1 and CNRS, F-69007 France}

\author{Eberhard Bodenschatz}
\affiliation{Max Planck Institute for Dynamics and Self-Organization (MPIDS), G\"ottingen, D-37077 Germany}
\affiliation{Institute for Nonlinear Dynamics, University of G\"ottingen, D-37077 Germany}
\affiliation{Laboratory of Atomic and Solid State Physics and Sibley School of Mechanical and Aerospace Engineering, Cornell University, Ithaca, NY 14853, USA}

\author{Haitao Xu}
\affiliation{Max Planck Institute for Dynamics and Self-Organization (MPIDS), G\"ottingen, D-37077 Germany}

\date{\today}

\begin{abstract}

{We describe the structure and dynamics of turbulence by the scale-dependent 
perceived velocity gradient tensor as supported by following four tracers, 
\textit{i.e.} fluid particles, that initially form a regular tetrahedron. 
We report results from experiments 
in a von K\'arm\'an swirling water flow and from numerical simulations of the 
incompressible Navier-Stokes equation. 
We analyze the statistics and the dynamics of 
the perceived rate of strain tensor and vorticity for initially regular tetrahedron of size $r_0$ from the dissipative to the integral scale. 
Just as for the true velocity gradient, at any instant, the perceived vorticity is also preferentially aligned with the intermediate eigenvector of the perceived rate of strain. However, in the perceived rate of strain eigenframe fixed at a given time $t=0$, the perceived vorticity evolves in time such as to align with the strongest eigendirection at $t=0$. This also applies to the true velocity gradient.
The experimental data at the higher Reynolds number 
suggests the existence of a self-similar regime in the inertial range. 
In particular, the dynamics of alignment of the perceived vorticity and strain 
can be rescaled by $t_0$, the turbulence time scale of the flow when the 
scale $r_0$ is in the inertial range.
For smaller Reynolds numbers we found the dynamics to be scale dependent.
}

\end{abstract}

\maketitle

\section{Introduction}
\label{sec:intro}

The fundamental question of how small scales are generated in turbulent 
flows is not fully understood~\cite{taylor:1938}. 
From 
a mathematical point of view, small scales are produced by the 
temporal amplification of the velocity gradients
when following a fluid element~\cite{pumir:1990,pauls:2006,gibbon:2008}. 
While the velocity gradient tensor is in principle available in direct 
numerical simulations (DNS), the rare, very intense fluctuations of the 
gradients make 
it numerically very challenging to accurately determine their statistical 
properties~\cite{schumacher:2007}. 
In laboratory flows, of course  the physics guarantees that all scales are 
resolved. However, 
measurements of the full velocity gradient tensor are in practice 
extremely 
difficult~\cite{Tsi92,vanderbos:2002,zeff:2003,luethi:2005,Wallace09,Wallace10}.

To address the question of the generation of small scales, one needs to know the evolution of the 
flow when following a fluid element. 
The recent development of Lagrangian measurement technology now allows {one}
to experimentally investigate this issue \cite{luethi:2005}. 
Nevertheless, the precise measurement of the velocity {{\it gradient} }
tensor at high Reynolds numbers will for the foreseeable future remain 
immensely
challenging due to temporal and spatial resolution constraints.

Because of the difficulty in determining reliably the velocity gradient,
a fruitful alternative approach consists in investigating a
coarse-grained 
version of the velocity tensor over a region of size $r_0$. In addition
to the possibility of determining the flow evolution from a Lagrangian point 
of view, and thus of addressing fundamental questions about the 
generation of small scales, this
approach enables 
the study of the scale dependence of flow properties, especially
in the inertial range of scales where the dynamics is believed to be 
determined by inertia only, and dissipation can be considered as
negligible. Several approaches can be used to define a coarse-grained 
velocity gradient tensor. 
One method consists in measuring the true 
coarse-grained tensor by averaging the velocity gradients over a well 
resolved region~\cite{vanderbos:2002,buxton:2011}.
The alternative strategy followed in this work rests on a reduced 
description~\cite{CPS99,XPB11}, 
based on defining a ``perceived velocity gradient tensor'' (denoted from now 
on by $\mathbf{M}(t)$ or $M_{ij}(t)$) that is supported only by four 
fluid elements, initially separated by a distance $r_0$ from each other, 
thus forming a regular tetrahedron. 
The tensor $\mathbf{M}(t)$ reduces to the true velocity gradient tensor 
$\mathbf{m}(t)$ when $r_0 \rightarrow 0$, or more precisely, when the scale 
$r_0$ is much smaller than the Kolmogorov scale, $\eta$, the smallest 
length scale in the flow. 
One physically important feature of this description is that
it provides insight not only on $\mathbf{M}$, but also on the geometry, 
\ie, the shape of the tetrahedron, and on its time evolution.
As we see later, the two aspects are strongly coupled \cite{CPS99,XPB11}.
Thus, the approach based on the perceived velocity gradient tensor and 
its evolution significantly differs from the one based on the true 
coarse-grained velocity gradient~\cite{Luthi07}.
>From a theoretical point of view, the description used here replaces
a continuous field by a relatively small number of degrees of freedom. 
It is one of the
hopes of the present approach that such a conceptual simplification could
help in devising simple tractable models, which would help shed new light on
turbulence.

The velocity gradient tensor has been shown to possess the property that instantaneously vorticity aligns with the {\it intermediate} stretching direction (eigenvector) of the rate of strain tensor \cite{siggia:1981b,Ashurst87,Tsi92,Meneveau11}, whose eigenvalue is predominantly positive.  This was surprising as vorticity was expected to align with the {\it strongest} stretching direction. Already  Taylor conjectured \cite{taylor:1937,taylor:1938} that the turbulent 
cascade mechanism relies on the amplification of vortices of scale $r_0$ by stretching that  subsequently leads to breakup of the vortices into smaller ones.
Our recent results~\cite{XPB11} provide new insight on this conjecture.
By analyzing the time evolution of tetrahedra of size $r_0$, we found 
that the perceived vorticity tends to align with the earlier direction of the {\it strongest} stretching, corresponding to the largest eigenvalue of the perceived rate of strain tensor.  For $r_0$ in the inertial range, this evolution depended only on $t/t_0$, where $t_0 \equiv (r_0^2/\dissip)^{1/3}$ is the time scale of eddies of size $r_0$ in a turbulent flow with energy dissipation rate $\dissip$.  From here on, we use ``vorticity'' and ''rate of strain'' both for the perceived as well as real quantities, the difference can be deduced from the scales given.

Here, we analyze the structure and the dynamics of $\mathbf{M}$, and 
its dependence on scale, thus extending our previous 
investigations~\cite{XPB11}.
Our experimental results at $R_\lambda = 350$ demonstrate 
that the dynamics
of the alignment of vorticity with the eigenvectors of the 
rate of strain, measured at the time where the tetrahedron is regular, depends on $r_0$
through the rescaled time $t/t_0$. This suggests
the existence of a self-similar regime in the inertial range.
At smaller Reynods numbers, we found the dynamics to be scale dependent.
The numerical results show that when $r_0 \rightarrow 0$ 
the properties of $\mathbf{M}$ extrapolate smoothly to those of the true 
velocity gradient tensor, and the dynamics of alignment between vorticity and strain is qualitatively similar with one notable difference: the characteristic time scale for tetrahedra with size $r_0 \lesssim \eta$ is of the order of $\tau_K$, the Kolmorogov time scale.

We begin, Section~\ref{sec:definition}, by describing the construction of 
the velocity gradient tensor $\mathbf{M}$. Our experimental 
and numerical techniques are explained in Section~\ref{sec:methods}. 
Section~\ref{sec:deformation} then presents our results concerning the 
evolution of the shape of the tetrahedra. In Section~\ref{sec:scale_PVGT}, 
we discuss the instantaneous correlations between strain and vorticity at any instant, and their dependence on the scale $r_0$. 
The dynamics of $\mathbf{M}$, in particular the
alignment between vorticity and the initial eigenvectors of strain 
are presented in Section~\ref{sec:dynamics}, which includes a subsection of simplified theoretical analysis of the alignment process at short times. 
Finally, Section~\ref{sec:discussion}  contains our concluding remarks.

\section{Definition of the perceived velocity gradient tensor $\mathbf{M}$}
\label{sec:definition}

Given fluid velocities at four points in a flow, we compute the perceived velocity gradient $\mathbf{M}$ as follows \cite{CPS99,XPB11}. 
Denoting the positions of the four points as ${\bf x}^a(t)$, and their velocities as ${\bf u}^a(t)$, $a = 1,..., 4$, 
we express the positions and velocities with respect to the center of
mass as ${\bf x}'^a = {\bf x}^a - {\bf x}^0$, and ${\bf u}'^a =
{\bf u}^a - {\bf u}^0$, where 
${\bf x}^0 = \frac{1}{4} ({\bf x}^1 + {\bf x}^2 + {\bf x}^3 + {\bf x}^4 )$
and
${\bf u}^0 = \frac{1}{4} ({\bf u}^1 + {\bf u}^2 + {\bf u}^3 + {\bf u}^4 )$.
In full generality, $M_{ij}$ is defined by minimizing the
quantity:
\begin{equation}
K = \sum_{a=1}^4 \sum_{i=1}^3 \Bigl({u'}_i^a - \sum_{j=1}^3 {x'}_j^a M_{ji} \Bigr)^2 .
\label{cost_function}
\end{equation}
In the equation defining the quantity to be minimized, $K$, the indices 
$a$ ($=1,..., 4$) refer to the index of the point, and $i$, $j$ ($ = 1,..., 3$)
to the component of the vector. 
By differentiating with respect to $M_{ij}$, one finds immediately that the 
condition to minimize $K$ is:
\begin{equation}
\sum_{k=1}^3 g_{ik} M_{kj} = W_{ij} ,
\label{eq_min}
\end{equation}
where the matrices $\mathbf{g}$ and $\mathbf{W}$ are defined by:
\begin{equation}
g_{ij} = \sum_{a=1}^4 {x'}_i^a {x'}_j^a ~~~ {\rm and } ~~~
W_{ij} = \sum_{a=1}^4 {x'}_i^a {u'}_j^a .
\label{def_g_W}
\end{equation}
Under the condition that $\mathbf{g}$ is invertible, the matrix 
$\mathbf{M}$ defined by:
\begin{equation}
\mathbf{M} = \mathbf{g}^{-1} \mathbf{W} ,
\label{expr_M}
\end{equation}
provides the
best fit approximation of $\mathbf{M}$ based on four points.

As the flow is incompressible, we impose the constraint that $\textrm{tr}(\mathbf{M}) =0$.
This can be done by determining $\mathbf{M}$, as explained in the previous 
paragraph, and  subtracting $\frac{1}{3} \textrm{tr}(\mathbf{M})\delta_{ij}$.
Alternatively, returning to the minimization condition, Eq.~\eqref{cost_function}, the trace can be imposed by adding a Lagrange multiplier, $\xi \sum_i M_{ii}$ to Eq.~\eqref{cost_function}. 
The two definitions, although technically slightly different, lead to
equal statistical properties, and to identical physical conclusions.

The construction used here, based on Eq.~\eqref{expr_M}, explicitly
requires the matrix $\mathbf{g}$ to be invertible. As a symmetric matrix, 
$\mathbf{g}$ is diagonalisable. Its eigenvalues $g_i$ are positive, and 
$\sqrt{g_i}$ characterize the extent of the tetrahedra in the $i^{th}$
eigendirection of $\mathbf{g}$. The matrix $\mathbf{g}$ becomes non-invertible 
when one of its eigenvalue tends to zero.
Thus, the condition that $\mathbf{g}$ is invertible
means physically that the four points of the tetrahedron are not coplanar.
The technical difficulty associated with the minimization of the quantity $K$ 
for highly flattened tetrahedra can be solved using standard Singular Value Decomposition algorithms \cite{NumRec}. 

In this work, we followed tetrahedra that are initially regular 
(close to regular in experiments), so as to identify flow properties
at a single scale.
The alignment dynamics studied here occurs {\it before} the tetrahedra become  strongly deformed \cite{XOB08,HYS11} or we did not use the highly deformed tetrahedra in the analysis. As shown in Section~\ref{sec:deformation}  the fraction of highly-flattened tetrahedra is negligible up to $t_0/4$ and only approximately 10\%  are highly-deformed at $t_0/2$. 
As will be shown in later chapters, the most interesting dynamics occur before $t_0/2$. To further reduce the effect of tetrahedron deformation on the accuracy of determination of $\mathbf{M}$, we excluded highly deformed tetrahedra from our statistics of $\mathbf{M}$ (see details in Section~\ref{sec:deformation}). This comes at the cost that the number of samples in our statistics is slightly reduced.

Similar to the treatment of the true velocity gradient, it is convenient to decompose $\mathbf{M}$ as the sum of a symmetric, 
strain-like part, $\mathbf{S}$, and an antisymmetric, vorticity-like part, 
$\mathbf{\Omega}$:
\begin{equation}
\mathbf{S} = \frac{1}{2} ( \mathbf{M} + \mathbf{M}^T) ~; ~~~ \mathbf{\Omega} = \frac{1}{2} ( \mathbf{M} - \mathbf{M}^T ) .
\label{def_str_vor}
\end{equation}
The symmetric matrix $\mathbf{S}$ describes the local straining motion. 
It is characterized by three 
real eigenvalues, $\strain_1$, $\strain_2$ and $\strain_3$, associated with three eigenvectors $\mathbf{e}_i$. 
In the following, the three eigenvalues are sorted in decreasing order: 
$\strain_1 \ge \strain_2 \ge \strain_3$. 
The antisymmetric matrix, $\mathbf{\Omega}$, describes the local rotation:
$\Omega_{ij} = \frac{1}{2} \epsilon_{ijk} \omega_k$, where $\epsilon_{ijk}$
is the completely antisymmetric tensor, so that
$\mathbf{\Omega} \cdot \mathbf{x} = \frac{1}{2} \mathbf{\omega} \cross \mathbf{x} $. 
In the following,
we characterize $\mathbf{\omega} $ by its norm $ | \omega | $ 
and its direction $ \mathbf{e}_\omega $, with 
$ | \mathbf{e}_\omega | = 1$. 
In the limit of very small tetrahedra ($r_0 \ll \eta$), the tensor $\mathbf{M}$ 
reduces to the true velocity gradient
tensor $\mathbf{m}$, defined by $m_{ij} \equiv \partial_j u_i$, and the 
vector $\mathbf{\omega}$ defined above reduces to the usual definition of vorticity, 
$\mathbf{\omega} \equiv \nabla \cross {\mathbf u}$.

\section{Experiments and numerical simulations}
\label{sec:methods}

\subsection{Experiments}
\label{subsec:exp}

Using image-based optical particle tracking, we measured experimentally the trajectories of tracer particles seeded in a swirling water flow between two counter-rotating baffled disks~\cite{laPorta01,Voth02}. 
{Measurements were done in the center of the apparatus, where the 
residence time of particles is the largest.}
The maximum Taylor-microscale Reynolds number is
$R_\lambda \approx 10^3$, where $R_\lambda \equiv \sqrt{15u'^4/\nu\dissip}$ 
with $u' \equiv \sqrt{ \langle u_x^2 + u_y^2 + u_z^2 \rangle/3 } $ being the fluctuation velocity and $\nu$ the kinematic viscosity. The symbol $\langle \rangle$ refers to the average of the fluctuating quantities in the flow.
Measurements were carried out by using three high-speed CMOS cameras and high repetition rate frequency doubled Nd:YAG lasers. Our particle tracking algorithm \cite{OXB06} allowed us to follow simultaneously hundreds of particles in this intense turbulent flow. The measured trajectories, however, were frequently interrupted due to experimental artifacts such as the fluctuation of the illumination light intensity, the background and electronic noise. For multi-particle measurements, such as tetrahedra, it is very important to connect these interrupted trajectory segments in order to obtain better statistics at  long-times. We therefore re-connected the trajectories in position-velocity space \cite{Xu08}, which were then smoothed and differentiated to obtain instantaneous particle velocities \cite{MCB04}. 

In this paper, we focus on results from a measurement at $R_\lambda = 350$, at which the scale separation between the integral scale $L = u'^3 /\dissip$ and the dissipative scale $\eta = ( \nu^3 / \dissip)^{1/4}$ is $L / \eta \approx 800$. As shown previously~\cite{XOVB07}, the temporal and spatial resolutions are sufficient to accurately measure the second moments of acceleration statistics. 

>From experimentally measured particle trajectories, we first identified instantaneous configurations of four points, 
separated from each other by a nominal distance $r_0$, within a tolerance of $\pm 10 \%$.
Once identified, the trajectories of the tetrahedra were followed for as long as all four points remained in the field of view.
Due to the limitation of particle seeding density in experiments, the data reported are mainly for tetrahedra with initial sizes between $L/17 - L/5$, or $50 - 180 \eta$. In this range of scales, we followed $10^7 - 10^8$ tetrahedra. 
The small tolerance of inter-particle distances at the nominal values results in initially nearly isotropic tetrahedra \cite{XOB08}. 

{
The particle trajectories used in this work were measured in an observation volume of size approximately (2.5 cm)$^3$ located at
the center of the apparatus, where the flow has been documented to be 
very close to isotropic~\cite{Voth02}. Deviations from isotropy may
be estimated from the well-known second order velocity structure functions, 
defined as 
$D_{LL}(\mathbf{r}) = \langle [(\mathbf{u}(\mathbf{x}+\mathbf{r}) - \mathbf{u}(\mathbf{x}))\cdot (\mathbf{r}/r)]^2 \rangle$ and $D_{NN}(\mathbf{r}) = \{\langle [(\mathbf{u}(\mathbf{x}+\mathbf{r}) - \mathbf{u}(\mathbf{x}))]^2 \rangle - D_{LL}\}/2$, where $r = \vert\mathbf{r}\vert$ is the separation distance. For isotropic turbulence, $D_{LL}$ and $D_{NN}$ depend only on $r$ and they satisfy the following relations (see \textit{e.g.}, \cite{pope:2000}): $D_{NN}(r) / D_{LL}(r) = 2$ for $r \ll \eta$ and $D_{NN}(r) / D_{LL}(r) = 4/3$. For the von K\'arm\'an flow in our experiment, due to the axisymmetric geometry, the statistics could depend on the angle $\theta$ between the axis of rotation and
the separation vector $\mathbf{r}$.  
We found that for $40 \eta \leq r \leq L/3$, the ratio $(3/4)D_{NN}(r,\theta) / D_{LL}(r,\theta)$ 
varies as a function of $\theta$ in the range $0.8 -- 1.3$, but averages 
out to be very close to 1, the isotropic value,
when considering all possible orientations of the separation. This provides
evidence both that the dependence on the angle $\theta$ is weak, and that
isotropy is recovered when averaging out over all possible directions~\cite{Tanveer:99,Taylor:03}.
Further evidence that the anisotropies in our flow are not a concern
comes from the fact that most experimental statistics derived from 
tetrahedra are in good agreement with DNS results using perfectly isotropic tetrahedra, see also~\cite{XPB11}.} 
The small deviation from precise isotropy in initial shapes, however, did give rise to discrepancies for certain quantities sensitive to initial shape, which we will discuss in detail later.

\subsection{Direct Numerical Simulations}
\label{subsec:dns}

We used a spectral code to solve the Navier-Stokes equations for the (Eulerian) velocity field 
$ {\bf u} ({\bf x} ,t) $:
\begin{eqnarray}
\partial_t {\bf u} ({\bf x},t)  + ( {\bf u } \cdot \nabla ) {\bf u} ({\bf x}, t) & = & - \nabla p ({\bf x}, t) + \nu \nabla^2 {\bf u}({\bf x}, t) \label{NSu} \\
\nabla \cdot {\bf u} ( {\bf x}, t) & = & 0 \label{NSi}
\end{eqnarray}
Eqs. \eqref{NSu} and \eqref{NSi} are integrated in a periodic box (size $2 \pi$), with
up to $384^3$ modes.
Energy is injected into the flow at large scale by letting the 
low wavenumber modes $|k| \le K$ (here, $K = 1.5$) evolve according
to the Euler equations truncated to the shell $|k | \le K$. This
maintains a constant amount of energy at the largest 
{scales~\cite{Pumir94}}.
Because the simulated flow is highly isotropic, we define the turbulence parameters based on the $x$ component of velocity. The Reynolds number is 
\begin{equation}
R_\lambda \equiv \frac{\taylorscale u'}{\nu} ,
\label{Rlambda}
\end{equation}
where the Taylor micro-scale defined as
\begin{equation}
\taylorscale^2 = \frac{u'^2}{\langle (\partial_x u_x)^2 \rangle}, 
\label{lambda}
\end{equation}
in which $u' \equiv \langle u_x^2 \rangle^{1/2}$.
The Kolmogorov length scale is $\eta = ( \nu^3 / \dissip)^{1/4}$, where $\dissip = \nu \langle \partial_j u_i \partial_j u_i \rangle$.
The available resolution has allowed us to simulate reliably turbulent flows
up to a Reynolds number
$R_\lambda = 170$.
Quantitatively, the largest wavenumber faithfully
simulated, $k_{max}$, is such that the product 
$k_{max} \times \eta \approx 1.4$.

 Defining the integral length scale by 
$L = u'^3/\dissip$, 
the ratio $L/\eta$ is approximately $300$ at $R_\lambda = 170$. We note 
that $L$ defined here is roughly twice as large
as the correlation length scale of the Eulerian velocity field:
\begin{equation}
L_{corr} \equiv \frac{\pi}{2  u'^2 } \int_0^{\infty} k^{-1} E(k) dk \approx L/2 ,
\label{L_corr}
\end{equation}
where $E(k)$ is the energy spectrum.

With the Eulerian velocity field determined by solving numerically 
Eqs. \eqref{NSu} and \eqref{NSi}, we also followed the motion of 
tetrahedra, whose vertices are tracer fluid particles,
evolving according to:
\begin{equation}
\frac{d {\bf x} (t) }{dt} = {\bf u}( {\bf x}(t),t) \label{lagrangian} .
\end{equation}
To this end, the values of ${\bf u}$, known after each time step on a regular
grid of collocation points, are interpolated at the location ${\bf x}(t)$ of 
the Lagrangian tracers, using accurate third order schemes \cite{PopeYeung88}. 

To study the dynamics using tetrahedra, we initialized tetrahedra of particles mutually separated by a distance $r_0$, 
which we varied over the full range of available scales, $\eta \lesssim r_0 \lesssim L$.
The statistical properties at a given scale $r_0$ have been determined by 
following at the minimum $125,000$ tetrahedra.

{
\section{Deformation of the tetrahedra} 
\label{sec:deformation}

We consider here the deformation occuring over a short time scale,
while the process of alignment between the perceived vorticity and strain is
taking place.
To quantify the deformation of tetrahedra, it is convenient to introduce
the tensor $\mathbf{g}$, defined in Eq.~\eqref{def_g_W}, and its eigenvalues~\cite{PSC00,Biferale05,XOB08,HYS11,XPB11}.
The trace of $\mathbf{g}$, which is the square of the radius of gyration of the four points,
quantifies the overall size of the object. The tensor $\mathbf{I} \equiv \mathbf{g}/\textrm{tr}(\mathbf{g})$, whose trace is $1$, characterizes the shape of the object. 
Specifically, because $\mathbf{I}$ is a symmetric matrix, it can be diagonalized in an orthogonal basis formed by its eigenvectors $\mathbf{e}_{Ii}$. 
In the following, we arrange the eigenvalues of $\mathbf{I}$ in 
decreasing order:
$I_1 \ge I_2 \ge I_3$. Since $I_1 + I_2 + I_3 = 1$, 
a regular tetrahedron corresponds to $I_1 = I_2 = I_3 = 1/3$, while a nearly co-planar configuration gives $I_3 \approx 0$. 

The matrix $\mathbf{g}$ is related to the moment of inertia tensor, $\mathbf{J}$, familiar in the context of mechanics of solid bodies, by \cite{XPB11}: 
\begin{equation}
J_{ij} \equiv  \textrm{tr}(\mathbf{g}) \delta_{ij} - g_{ij} .
\label{rel_g_J}
\end{equation}

\subsection{Alignment of tetrahedron geometry with the eigenvectors of the rate of strain}
\label{sec:alignment_geometry_strain}

The dynamics of tetrahedra is strongly influenced by the local rate of strain. 
In particular, it has been demonstrated \cite{XPB11} that the stretching in 
the direction of the 
largest eigenvalue of the strain, $\mathbf{e}_1$, leads to an elongation of
the tetrahedron in this direction, and, by conservation of angular momentum,
to an alignment of the vorticity $\mathbf{\omega}(t)$ in the direction $\mathbf{e}_1(0)$.
The results presented here generalize this observation: at short times, the 
deformation of the tetrahedra is very well aligned with the eigendirection
of the eigenvectors of the rate of strain tensor.

%
%
\begin{figure}
\begin{center}
\subfigure[]{
	\includegraphics[width=0.47\textwidth,angle=0]{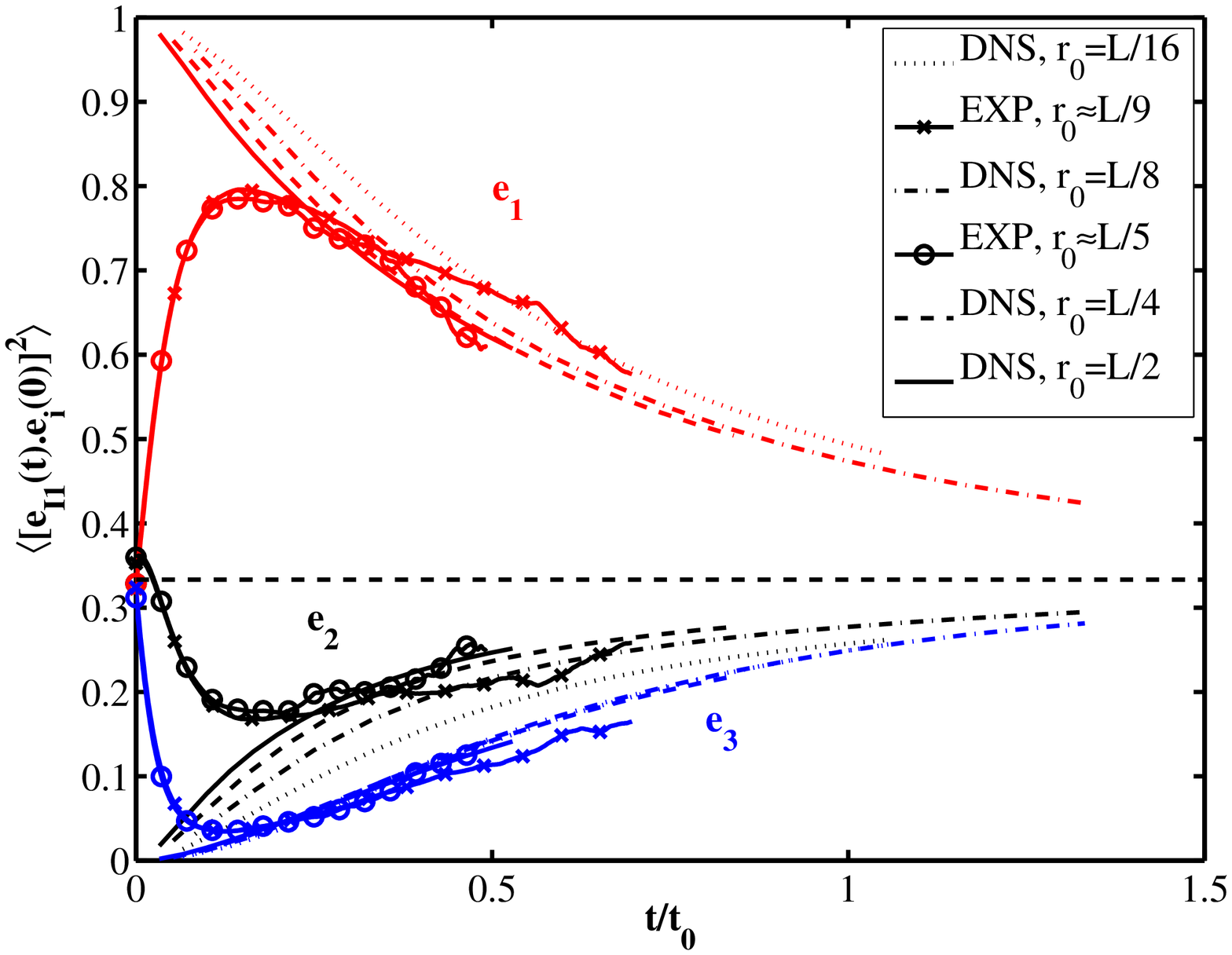}
}
\subfigure[]{
	\includegraphics[width=0.47\textwidth,angle=0]{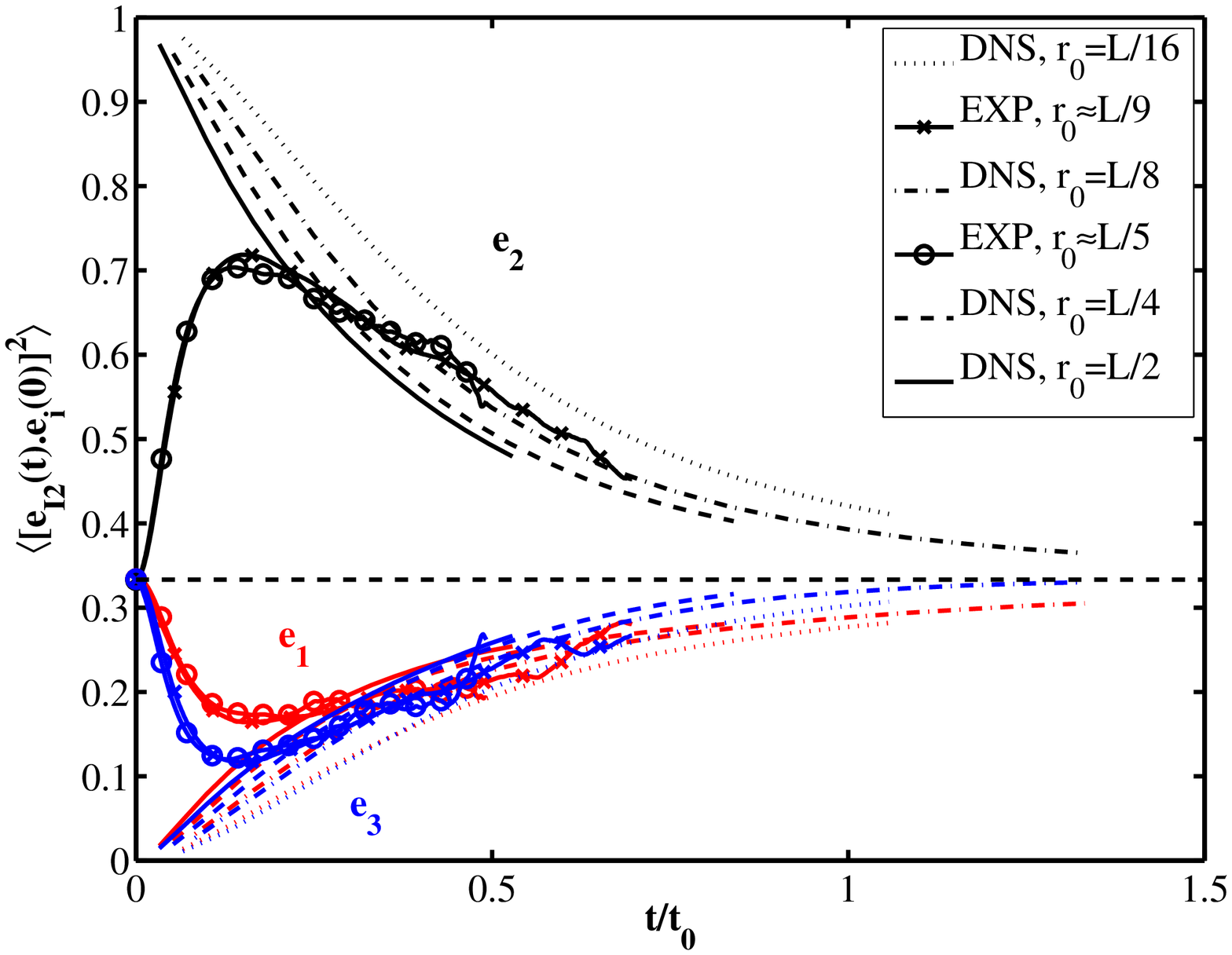}
}
\subfigure[]{
	\includegraphics[width=0.47\textwidth,angle=0]{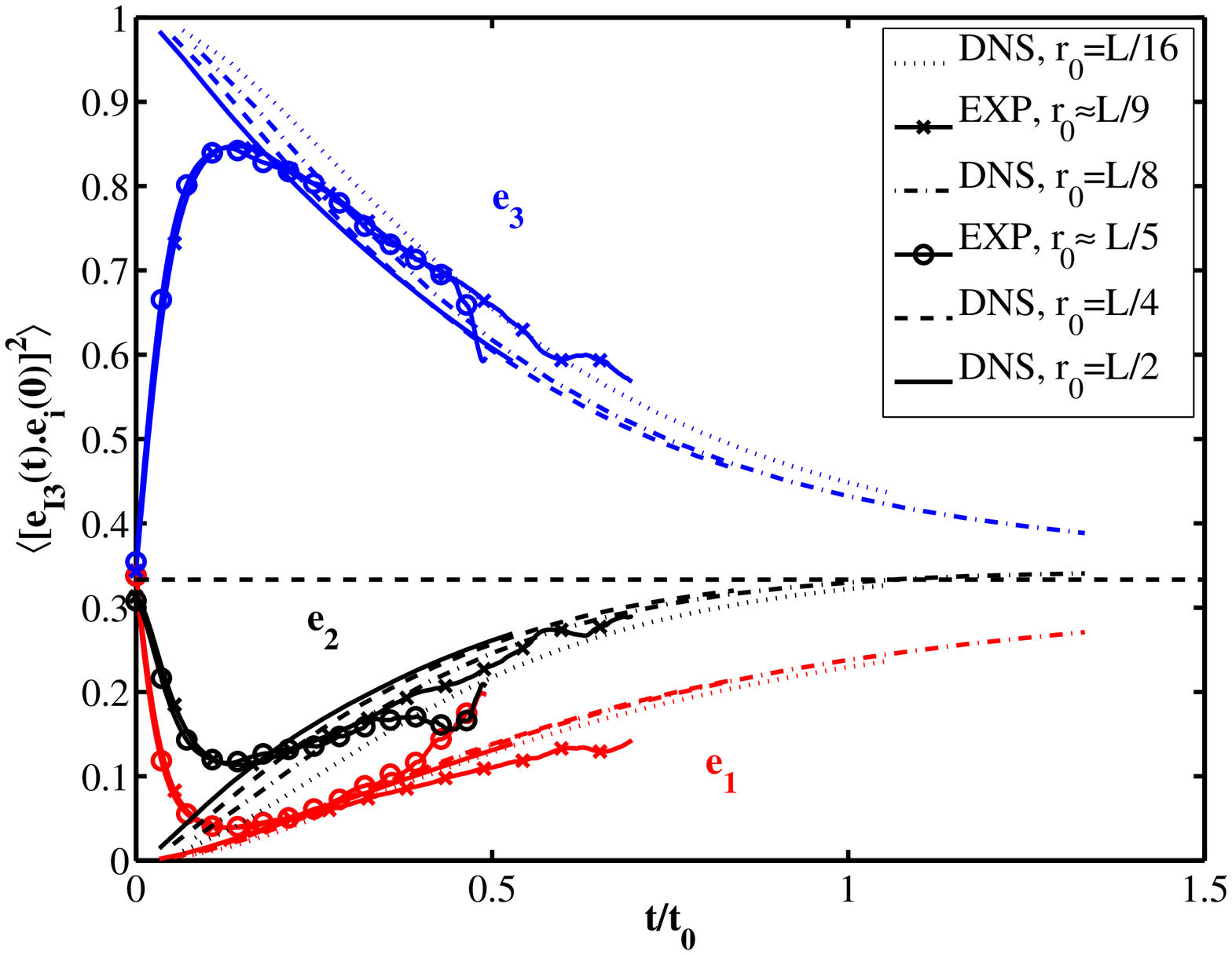}
}
\caption{(Color online)
Deformation of tetrahedra: alignment of the principal axes of the tetrahedra with
the eigenvectors of the strain. The ensemble averages
$\langle [\mathbf{e}_{Ii}(t) \cdot \mathbf{e}_j(0)]^2 \rangle $ as a function of
time for 
(a) $i=1$ (longest axis),
(b) $i=2$ (intermediate axis) and
(c) $i=3$ (shortest axis of the tetrahedron). 
As expected, the $i^{th}$ axis of the tetrahedron, $\mathbf{e}_{Ii}(t)$, aligns 
perfectly at very short times with the $i^{th}$ eigenvector of the strain, $\mathbf{e}_i(0)$. 
After a time of order $\sim t_0$, the alignment relaxes and 
the axes of the tetrahedron do not show any particular alignment with any of
the eigenvectors of the strain. 
The DNS data correspond to $R_\lambda 
= 170$, whereas the experiments to $R_\lambda = 350$.
}
\label{fig:deform_proj}
\end{center}
\end{figure}

Fig.~\ref{fig:deform_proj} shows how the principal axes of initially regular tetrahedra,
$\mathbf{e}_{Ii}(t)$,  align with the eigenvectors of the rate of strain, $\mathbf{e}_i (0)$. 
For initially isotropic tetrahedra, as in DNS, the matrix $\mathbf{I}$ is diagonal at $t=0$. 
A very small deformation of the tetrahedra due to the strain breaks the isotropy,
and leads to the immediate alignment of the principal axes of the tetrahedra with 
the strain eigenvectors. This is reflected in Fig.~\ref{fig:deform_proj} by the
perfect alignment between 
$\mathbf{e}_k(0)$ and $\mathbf{e}_{Ik}(t) $ at very small times $t$
($t = 0^+$). 
By $t = 0{^+}$, we refer to a strictly positive, but very small value of time. 
The tetrahedra studied experimentally are not exactly isotropic initially. Instead,
the eigenvalues of $\mathbf{I}$ are slightly different from each other:
$I_i = \frac{1}{3} + \delta I_i$, where $ | \delta I_i | \ll 1$.
At $t=0$, the corresponding three eigenvectors of $\mathbf{I}$ are 
uncorrelated with the strain eigenvectors, i.e., 
$\langle [\mathbf{e}_{Ik}(0) \cdot \mathbf{e}_j(0)]^2 \rangle \approx 1/3$.
This explains the difference at short times seen in Fig.~\ref{fig:deform_proj} between the curves obtained from 
DNS and from experiment.
On the other hand, under the action of the strain, the tetrahedra deform quickly and their principal axes align with the eigenvectors of the strain in a time of approximately $t_0/10$, after which the tetrahedron evolution is independent of initial shape.
We note that over the range of scales shown here,  $L/16 \le r_0 \le L/2$, the dynamics
of alignment is essentially self-similar, once expressed in units of $t_0$.
In a time of order $t_0$, the axes $\mathbf{e}_{Ik}(t)$ lose any memory of the initial alignment with $\mathbf{e}_j(0)$; 
all values of $\langle [\mathbf{e}_{Ik}(t) \cdot \mathbf{e}_j(0)]^2 \rangle$ relax towards $1/3$. 
This is consistent with the fact that $t_0$, which is the time introduced in Kolmogorov theory, is the proper correlation times of an object initially with a scale $r_0$. 
Small deviations are visible for DNS data of $r_0 = L/16 \approx 20 \eta$, which is known to be the lower limit of the inertial range.

\subsection{Flattening of the tetrahedra}
\label{sec:tetrad_flattening}

Along with the stretching of the tetrahedra in the direction $\mathbf{e}_1(0)$,
documented in particular in~\cite{XPB11}, one expects a strong compression
in the direction $\mathbf{e}_3(0)$, corresponding to the most negative
eigenvalue of the perceived strain. This has been shown to lead to very 
significantly flattened tetrads \cite{Biferale05,XOB08,HYS11}. In 
particular, the probability
distribution function (PDF) of $I_3$ is sharply peaked around $0$ 
at late times, $t/t_0 \gtrsim 1$. 
As explained in Section~\ref{sec:definition}, this can be the cause of
a serious limitation in our
construction of the perceived velocity gradient,
as implied by Eq.~\eqref{expr_M}. 
A good understanding of the dynamics leading to flattened tetrahedra is 
thus of interest not only for its own sake, but also in relation to
the dynamics of alignment between perceived vorticity and strain.
This subsection therefore discusses the flattening of tetrahedra,
and demonstrates that the formation of very flat configurations is not faster
than the alignment process between vorticity and strain.

%
%
\begin{figure}[t]
\begin{center}
\subfigure[]{
	\includegraphics[width=0.47\textwidth,angle=0]{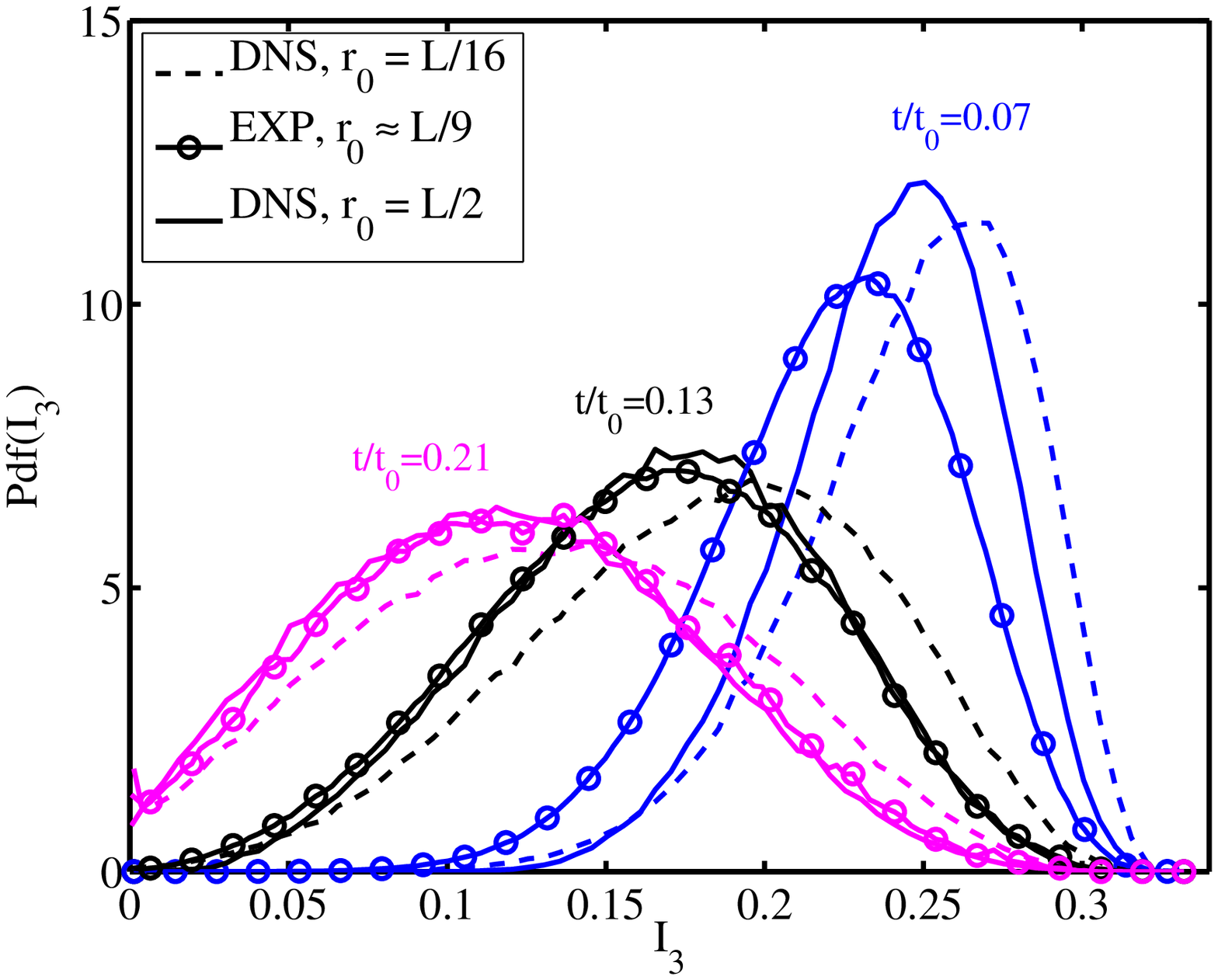}
}
\subfigure[]{
	\includegraphics[width=0.47\textwidth,angle=0]{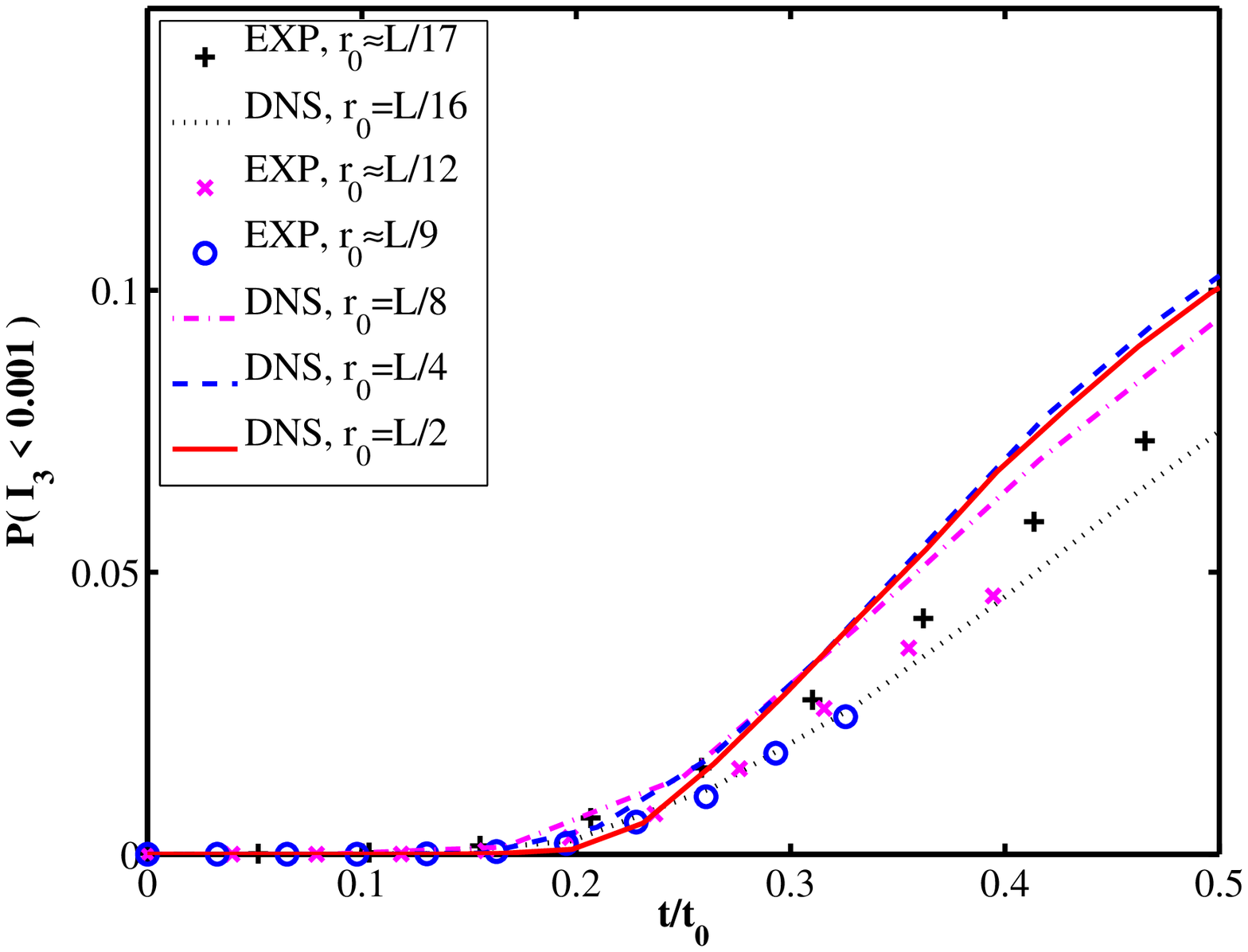}
}
\caption{(Color online) Flattening of tetrahedra. (a) PDF of $I_3$ at several 
times, $t/t_0 = 0.07 $, $0.13$ and $0.21$, for 
different initial tetrahedron sizes ($r_0$ from $L/16$ to $L/2$) in 
both DNS ($R_\lambda = 170$) and experiment ($R_\lambda = 350$). 
The observed peak of the PDF decreases when time increases. At $t = 0.21 t_0$, 
a non-zero PDF at $I_3 \approx 0$ is observed. 
The probability that $I_3 \le 10^{-3}$ is shown in (b) for several values of $r_0$, as a function of the normalized time $t/t_0$.
A steep growth of the probability $P(I_3 \le 10^{-3} )$ is seen for $t \gtrsim t_0/4$. 
The growth of the probability does not depend much on the scale $r_0$ over the inertial range: 
from $r_0 = L/16  \approx 20 \eta$ in DNS and $r_0 = L/17  \approx 50 \eta$ in experiments to $r_0 = L/2$.
}
\label{fig:small_I3}
\end{center}
\end{figure}

To this end, Fig.~\ref{fig:small_I3} shows the evolution of the PDF of $I_3$ 
(a), and the evolution of the probability that the value of $I_3$ is less 
than a very small number, $I_3 \le 10^{-3}$ (b).
Initially, the probability of $I_3$ is peaked near $I_3 = 1/3$. 
In DNS, which starts with precisely isotropic tetrahedra, the PDF is originally a $\delta$-function at $I_3 = 1/3$.
In experiments, where we select particles that form nearly isotropic tetrahedra, the initial PDF is not a $\delta$-function, but is still narrowly distributed around $I_3 \approx 0.3$, as shown before\cite{XOB08}.
As time increases, the peak of the PDF shifts towards
smaller values of $I_3$ (Fig.~\ref{fig:small_I3}a), reflecting a tendency
of the tetrahedra to flatten. The evolution of the PDF does not depend very
much on the initial size of the tetrahedron, as seen when comparing
DNS and experimental data ranging from $r_0 = L/16 \approx 20 \eta$ in DNS to $r_0 = L/2$. 
A noticeable fraction of the tetrahedra becomes flattened after $ t \gtrsim t_0/4$ (see Fig.~\ref{fig:small_I3}(b)), consistent
with the sharply peaked distribution of $I_3$ observed at later times \cite{Biferale05,XOB08,HYS11}. 
>From a flat configuration, with all points in the same plane, it is not
possible to extract any information on the variation of the velocity in 
the plane transverse to the plane.
As explained in Section~\ref{sec:definition} however,
this does not prevent us from computing $\mathbf{M}$ using Singular Value Decomposition. However, it points to a limitation on the accuracy of the procedure. Therefore, in the results reported here, we excluded tetrahedra with $I_3 \leq 10^{-3}$ from the statistics of $\mathbf{M}$. As seen from Fig.~\ref{fig:small_I3}(b), this leads to at maximum a 10\% reduction in the number of samples at time $t/t_0 = 0.5$, while most of the interesting dynamics occurs before this time.

The results of this section thus show a very strong alignment between the
axes of the tetrahedra and those of strain, see Fig.~\ref{fig:deform_proj},
and that the dynamics of
flattening of the tetrahedra happens over a time 
of the order of $t_0/4$, see Fig.~\ref{fig:small_I3}. In addition, we find
that the dynamics is essentially self-similar, 
insofar as the time-dependent correlations presented here 
are independent of scale, once time has been expressed in units of $t_0$.
This is consistent with earlier results \cite{XOB08} demonstrating that 
the evolution of the shape factors, $I_i(t)$, is also self-similar. 
The ensuing picture is thus that in a statistical sense, initially regular
tetrahedra of size $r_0$ evolves with a time scale $\sim (r_0^2/\dissip)^{1/3}$,
when $r_0$ in the inertial range.
}

\section{Properties of $\mathbf{M}$: Scale dependence (Instantaneous statistics)}
\label{sec:scale_PVGT}

Previous investigations~\cite{siggia:1981b,Ashurst87,She91,Tsi92,Wallace09} 
of the structure of the true velocity gradient tensor, 
$\mathbf{m}$, reveal two main properties of the instantaneous statistics of $\mathbf{m}$: 

{\it (i)} The direction of the vorticity $\mathbf{e}_\omega$ is preferentially aligned with $\mathbf{e}_2$, the direction of the intermediate eigenvalue of the rate of strain, $\mathbf{s} = \frac{1}{2} ( \mathbf{m} + \mathbf{m}^T)$,
but does not show any particular alignment with $\mathbf{e}_1$, the strongest eigen-direction of the rate of strain.

{\it (ii)} The intermediate eigenvalue of the rate of strain, $\strain_2$, is mostly positive, which implies that the product of the eigenvalues of $\mathbf{s}$ is negative and thus ensures vortex stretching $\langle \mathbf{\omega} \cdot
\mathbf{s} \cdot \mathbf{\omega} \rangle \ge 0$, an important property of
turbulent flows \cite{Betchov56}.

It is natural to ask what are the instantaneous properties of the perceived velocity gradient $\mathbf{M}$.

On general grounds, one expects that the properties of $\mathbf{M}$ will depend
on the size $r_0$. For $r_0 \lesssim \eta$, $M$ coincides with the true velocity
gradient tensor. As the separation $r_0$ between the points of the
tetrahedron increase, the velocities at the four points become less and less 
correlated. For $r_0 \gg L$, $\mathbf{M}$ is not expected to 
reflect any particular property of the turbulent dynamics.
On the other hand, one may expect that in the inertial range, some properties
become scale-independent, consistent with the notion that the 
flow shows some degree of self-similarity.

\subsection{Alignment between vorticity and strain}
\label{subsec:algnmt_vort_strn}

\begin{figure}
\begin{center}
\subfigure[]{
	\includegraphics[width=0.47\textwidth,angle=0]{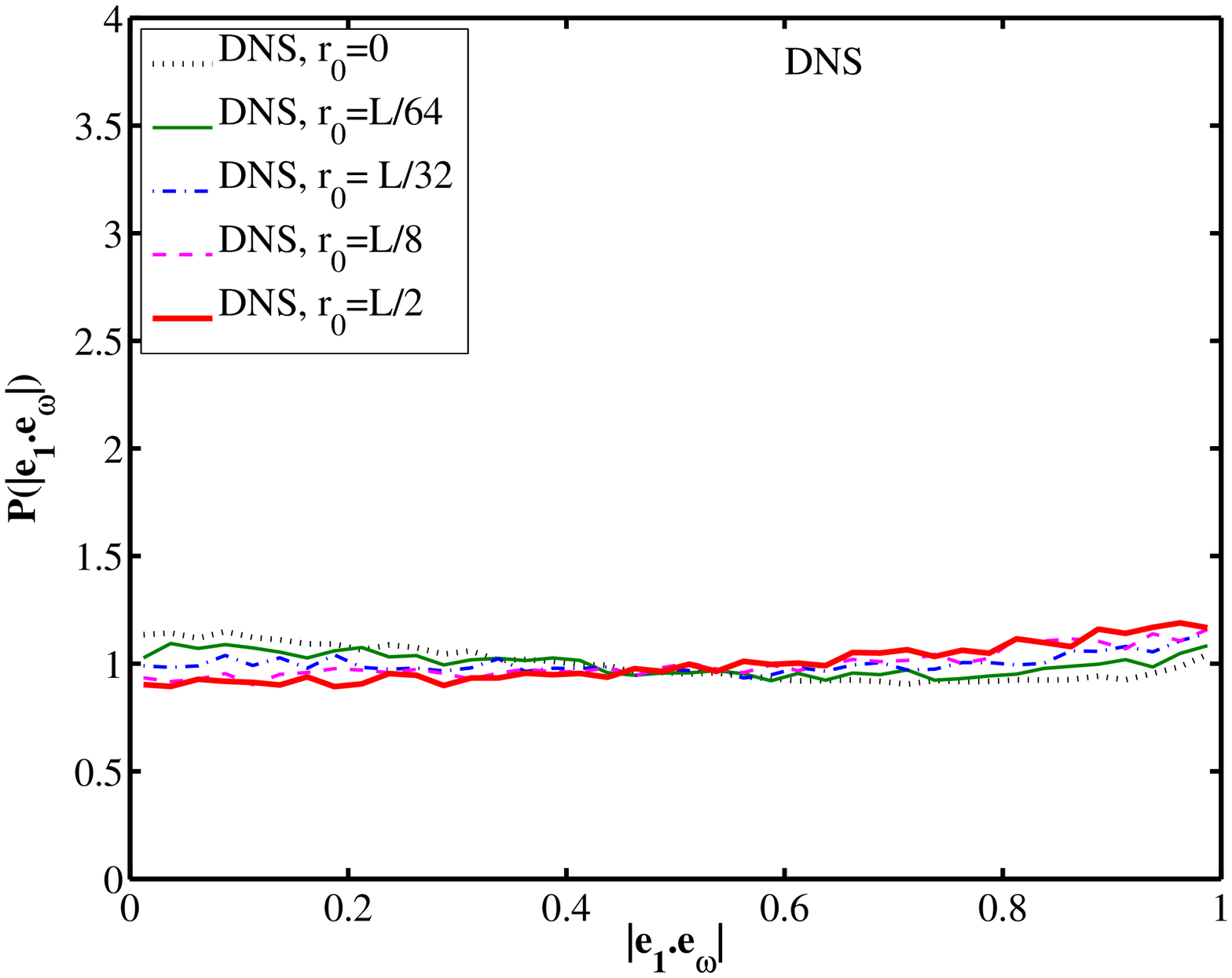}
}
\subfigure[]{
	\includegraphics[width=0.47\textwidth,angle=0]{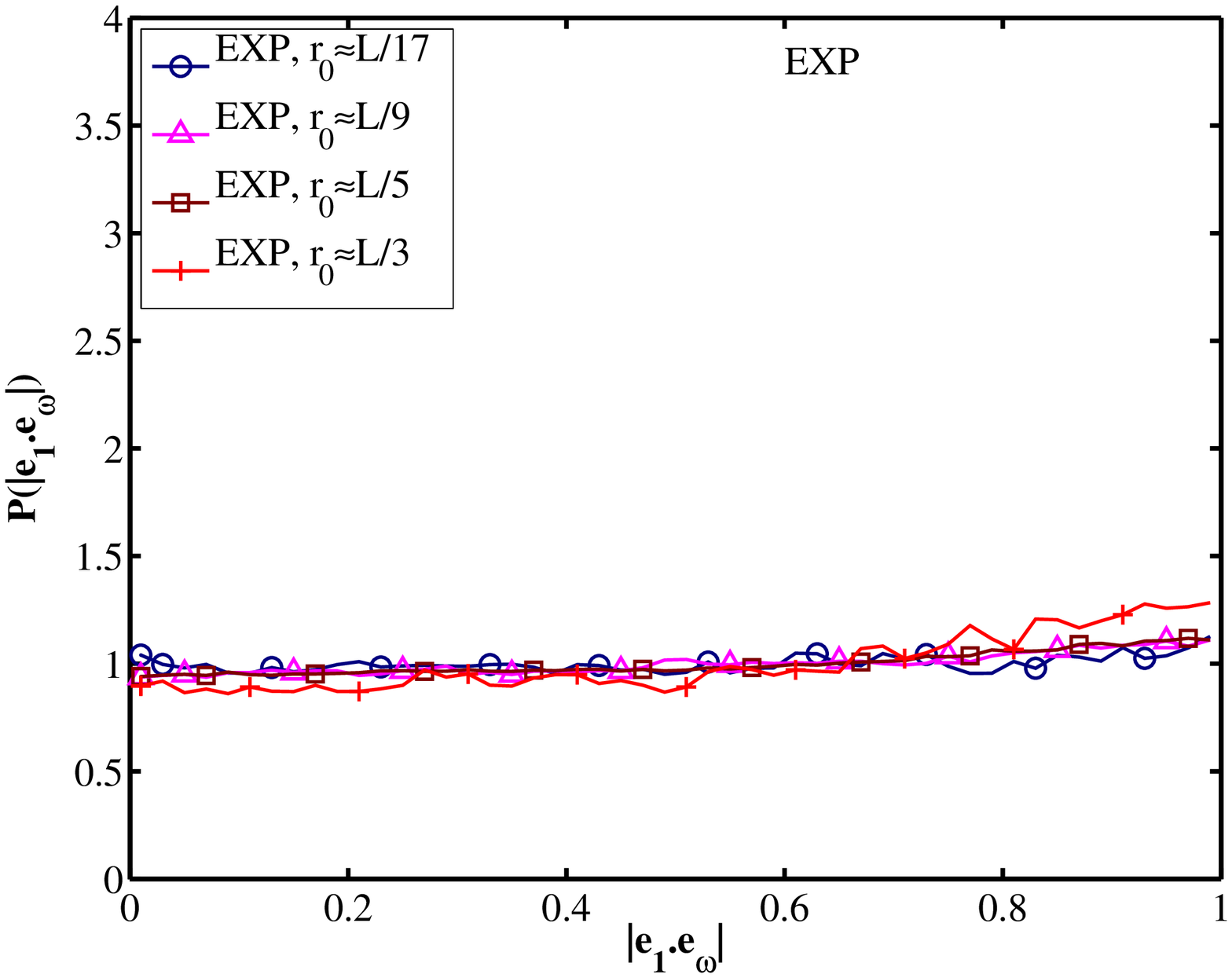} 
}
\\
\subfigure[]{
	\includegraphics[width=0.47\textwidth,angle=0]{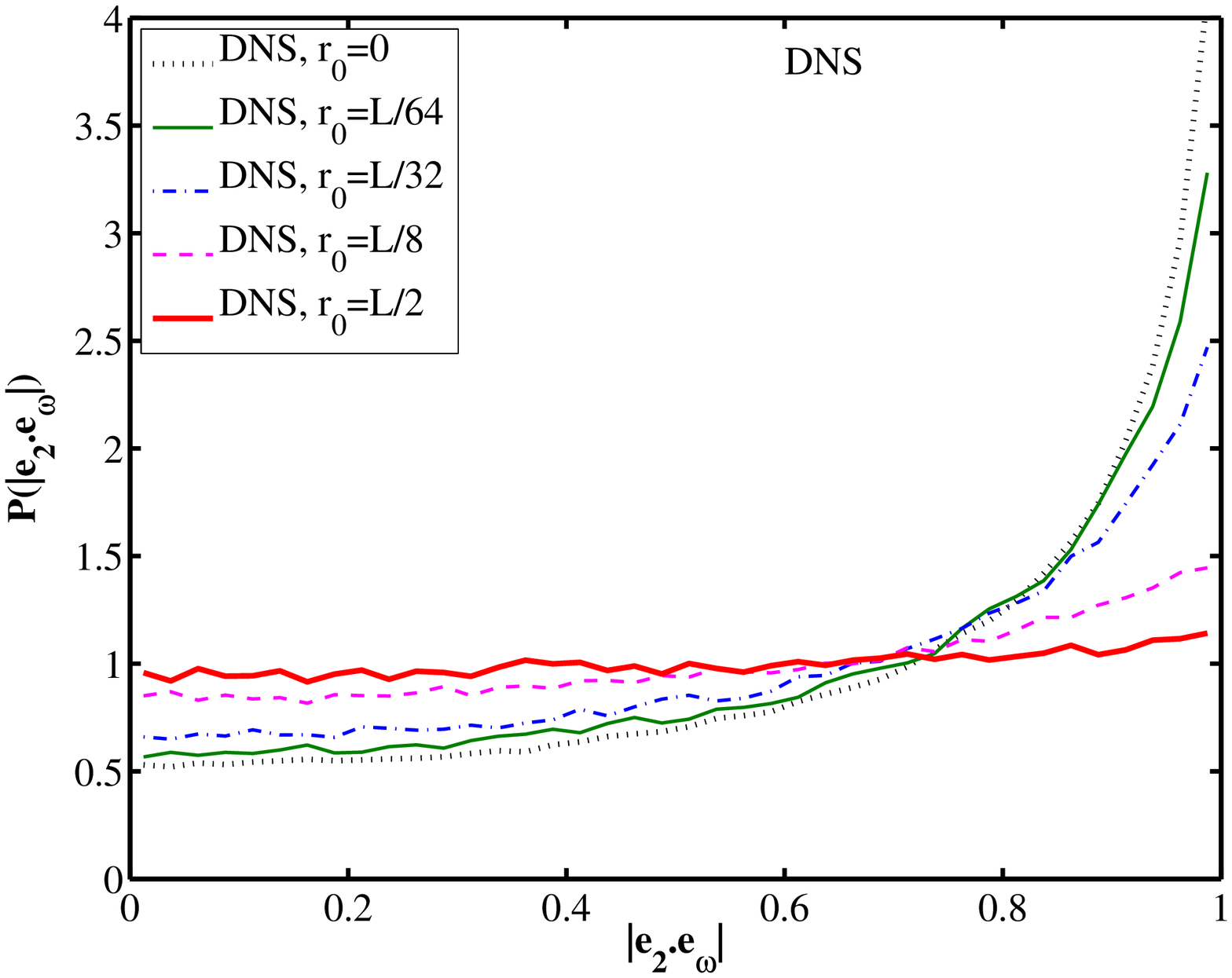}
}
\subfigure[]{
	\includegraphics[width=0.47\textwidth,angle=0]{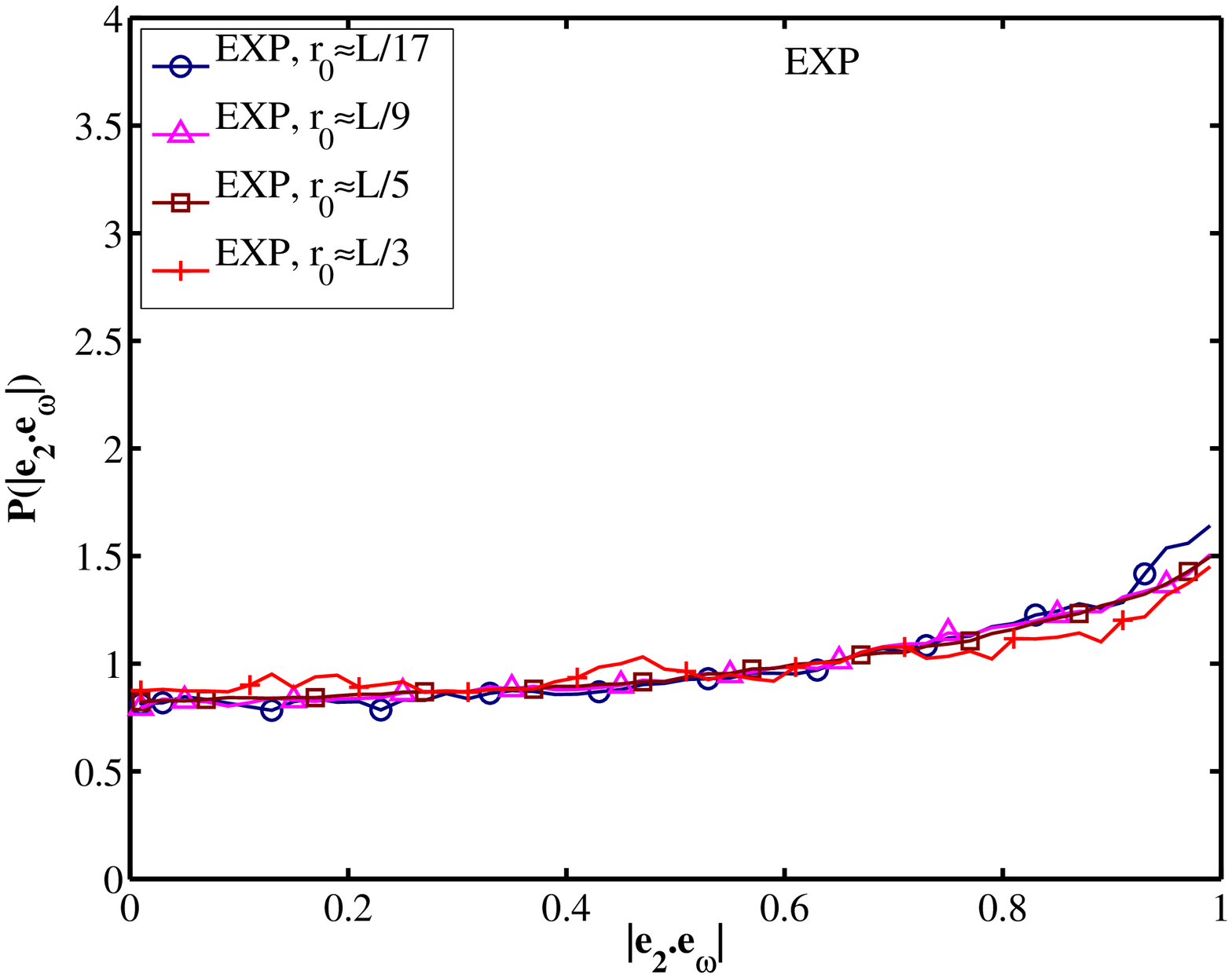}
}
\\
\subfigure[]{
	\includegraphics[width=0.47\textwidth,angle=0]{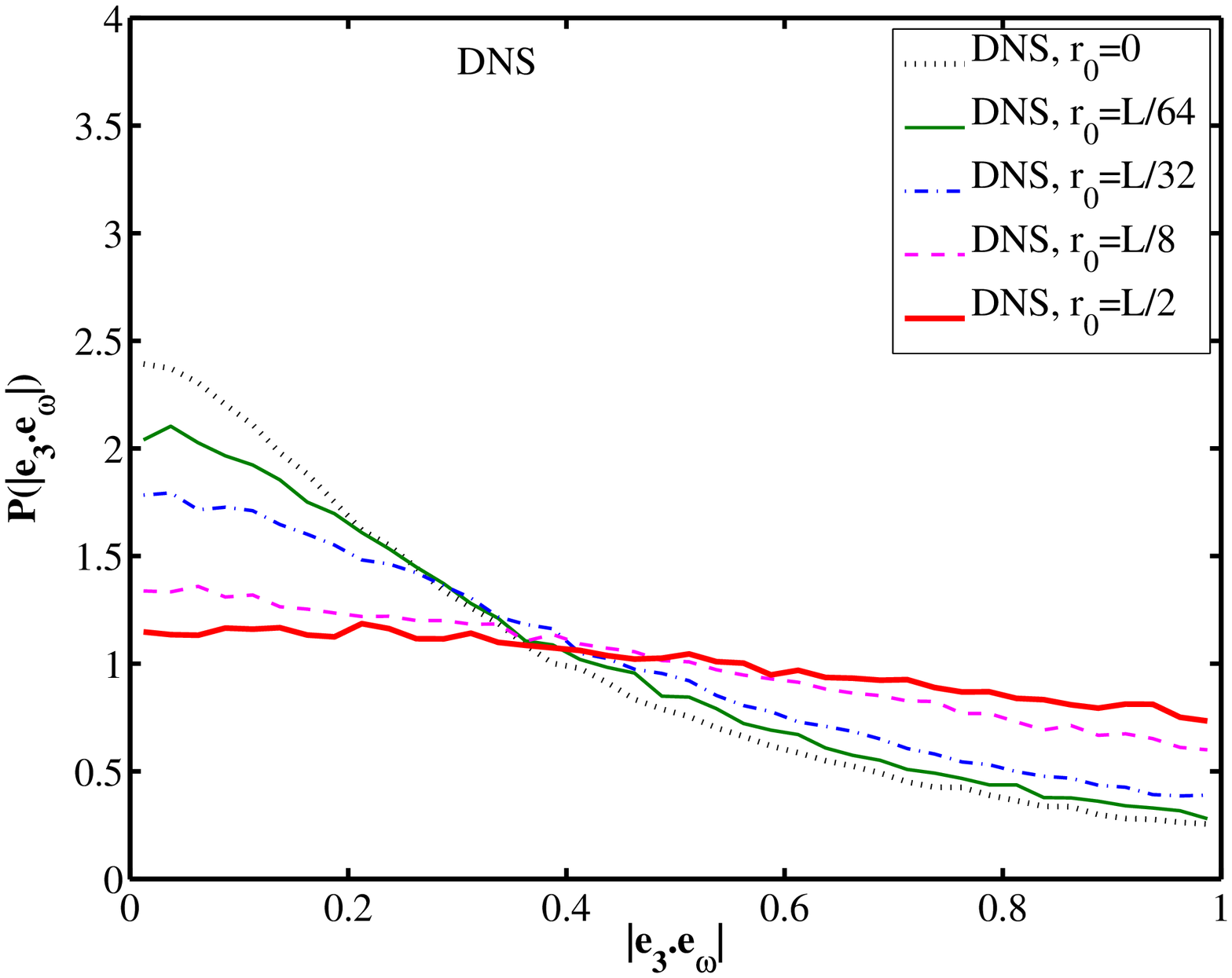} 
}
\subfigure[]{
	\includegraphics[width=0.47\textwidth,angle=0]{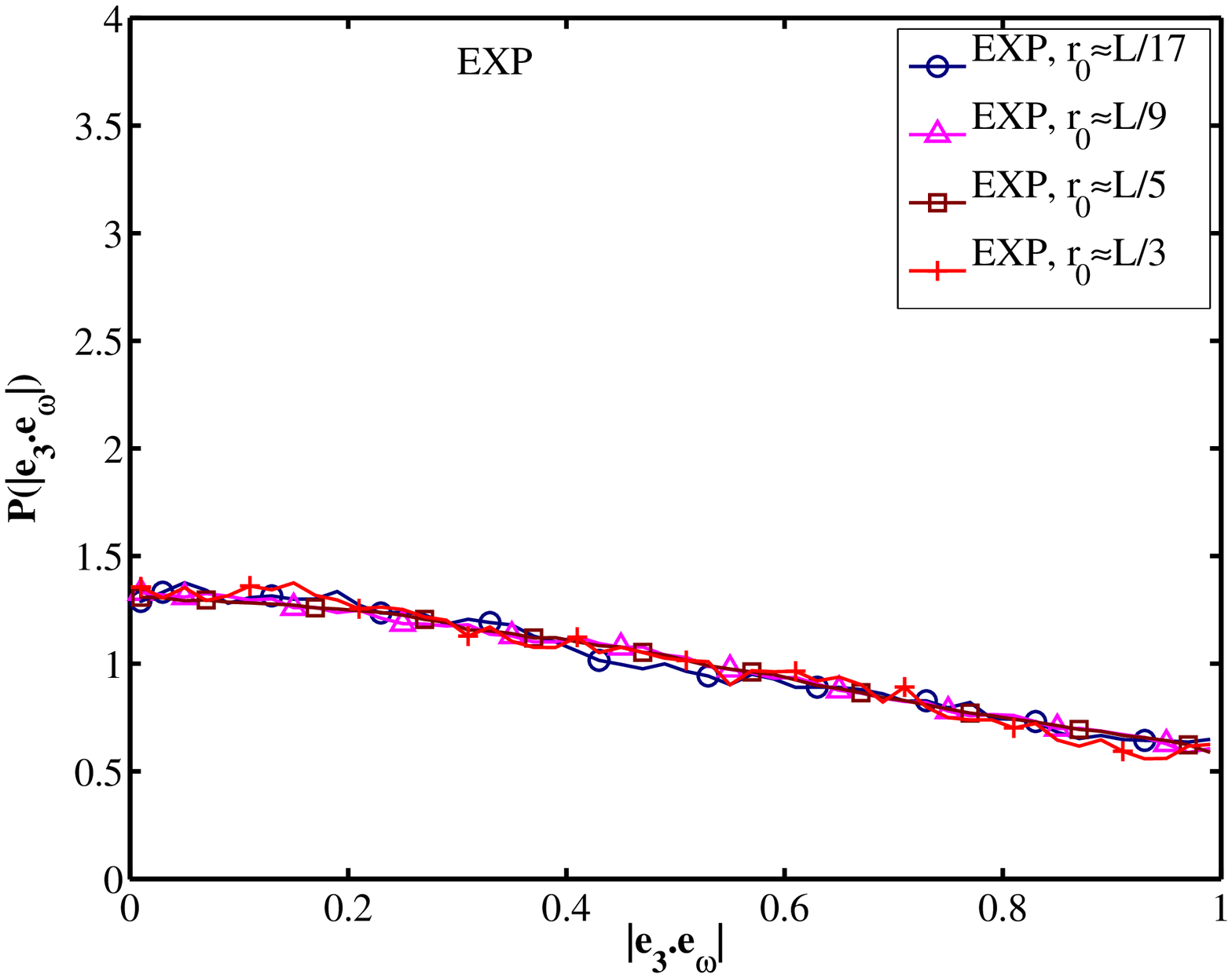}
}
\caption{(Color online) Scale dependence of the alignment of vorticity and 
strain {(instantaneous statistics)}: The PDFs of 
$|\mathbf{e}_i \cdot \mathbf{e}_\omega |$ from both DNS (panels a,c,e) and experiments (panels b,d,f)
for tetrahedra with different sizes $r_0$ from dissipative ($r_0 = 0$ in DNS) to integral ($L/2$) scales.
Panels (a) and (b) are for $i=1$; (c) and (d) for $i=2$; and (e) and (f) for $i=3$.
The lack of alignment of vorticity with $\mathbf{e}_1$ is nearly independent of scale.
The PDF of $|\mathbf{e}_{\omega} \cdot \mathbf{e}_2 |$ sharply peaks at $1$ at very small scales, in particular for the true velocity gradient tensor ($r_0 = 0$). The effect weakens as the scale increases.
Similarly, the maximum at $0$ of the PDF of $|\mathbf{e}_{\omega} \cdot \mathbf{e}_3 |$ at very small scales becomes milder as scale increases.
The DNS data correspond to $R_\lambda 
= 170$, whereas the experiments to $R_\lambda = 350$.
}
\label{fig:pdf_ei_eom}
\end{center}
\end{figure}

\begin{figure}
\begin{center}
\includegraphics[width=0.6\textwidth,angle=0]{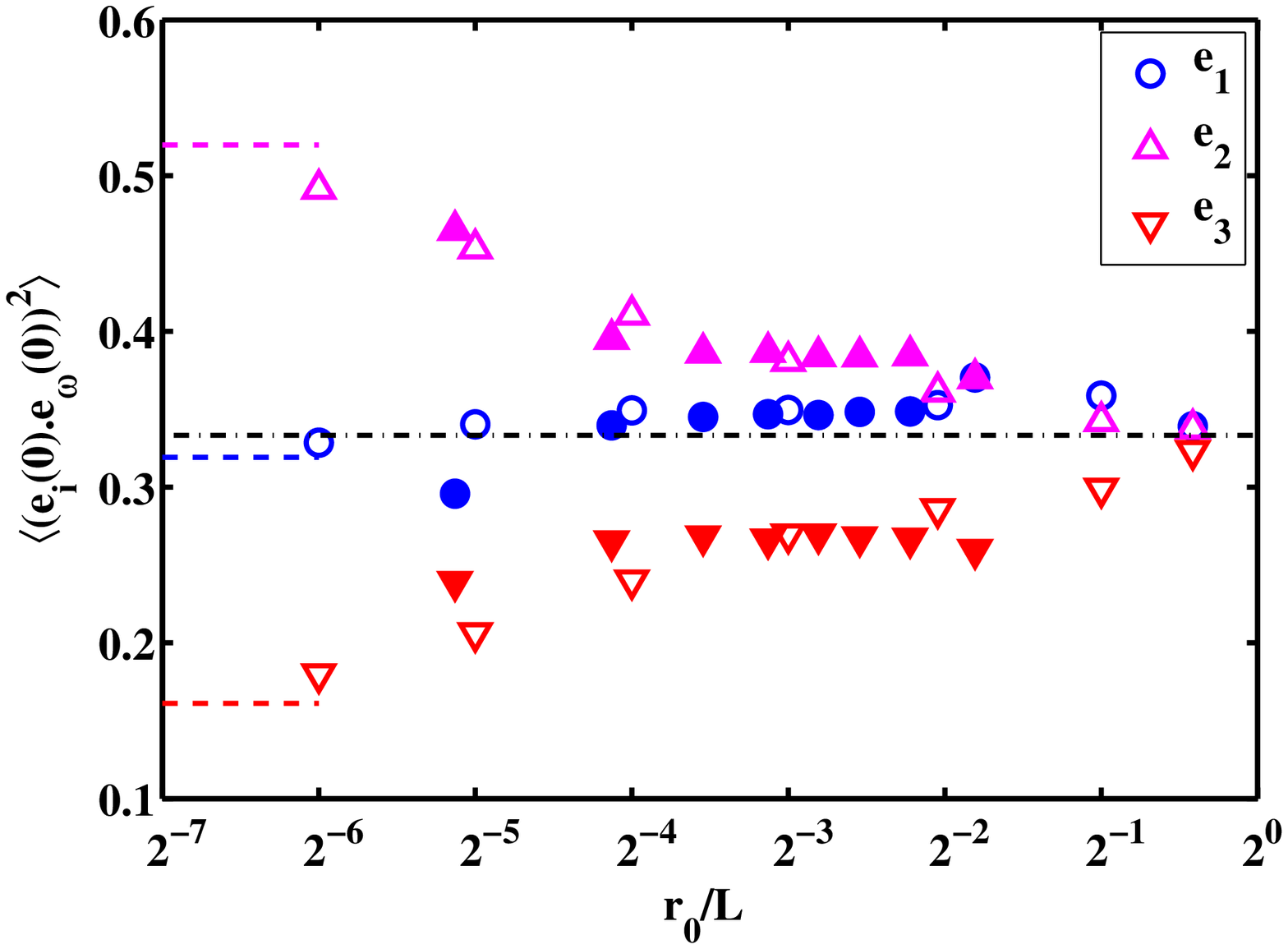}
\caption{(Color online)
The mean values $\langle ( \mathbf{e}_\omega(0) \cdot \mathbf{e}_i (0))^2 \rangle$. The open symbols are DNS data ($R_\lambda = 170$) and the filled symbols are experimental results ($R_\lambda = 350$).
}
\label{fig:mean_ei_eom}
\end{center}
\end{figure}

The PDFs of the cosines of the instantaneous angles  between $\mathbf{e}_\omega$ and $\mathbf{e}_i$, are shown in Fig.~\ref{fig:pdf_ei_eom}.
The PDF for $i=1$ (Fig.~\ref{fig:pdf_ei_eom}{{(a) and (b)}}) hardly varies through scale, and remains nearly uniform: 
$P(| \mathbf{e}_1 \cdot \mathbf{e}_\omega | ) \approx 1$, consistent with
previous observations with the true velocity gradient tensor \cite{Ashurst87,Tsi92}.
{In Fig.~\ref{fig:pdf_ei_eom}(c) and (e)}, the alignment corresponding to the true velocity gradient tensor $\mathbf{m}$ ({\it i.e.}, $r_0 = 0$) are the ones with the largest variations. 
As scale increases, the alignment between $\mathbf{e}_2$ and $\mathbf{e}_\omega$ becomes less pronounced, as shown by the 
observed decrease of the PDF at $| \mathbf{e}_2 \cdot \mathbf{e}_\omega | = 1$. 
Similarly, the tendency of $\mathbf{e}_\omega$ to be perpendicular to $\mathbf{e}_3$ diminishes when scale increases.
As the scale $r_0$ approaches $L$, the PDF becomes flat. 
This tendency can be clearly seen in Fig.~\ref{fig:mean_ei_eom}, which shows the
averaged values $\langle ( \mathbf{e}_i \cdot \mathbf{e}_\omega )^2 \rangle$.
When $r_0$ increases from dissipative to integral scales, 
$ \langle ( \mathbf{e}_\omega \cdot \mathbf{e}_2)^2 \rangle$ continuously 
decreases from the larger magnitute for the true velocity gradient tensor
(shown by the horizontal dashed lines in Fig.~\ref{fig:mean_ei_eom}) towards 
$1/3$, which corresponds to two randomly selected vectors.
In the same spirit, $ \langle ( \mathbf{e}_\omega \cdot \mathbf{e}_3)^2 \rangle$
continuously increases from the lower magnitude of the velocity gradient tensor to 
$1/3$ when $r_0$ increases.
In Fig.~\ref{fig:mean_ei_eom} are shown both the numerical data 
($R_\lambda = 170$, open symbols) and the experimental data ($R_\lambda = 350$,
filled symbols). Remarkably, 
{in both Figs.~\ref{fig:pdf_ei_eom} and \ref{fig:mean_ei_eom},} 
the lower Reynolds number data show a 
continuous variation of the {statistics of the} cosines 
$\mathbf{e}_2 \cdot \mathbf{e}_\omega$ and $\mathbf{e}_3 \cdot \mathbf{e}_\omega$
with scale. The variation is  
particular strong when $r_0$ decreases towards the dissipative range,
{\it i.e.}, for $r_0 \lesssim L/16 \approx 20 \eta$,
In comparison, the higher Reynolds number data suggest that over the accessible range of scales,
the properties of the flow are essentially scale independent. 
This suggests that at sufficiently large Reynolds numbers, the dynamics of $\mathbf{M}$
may show some self-similarity, as far as the relation between vorticity and
strain is concerned.
This remains to be studied further with data at larger Reynolds numbers.

%
%

%
%

The results summarized in Figs.~\ref{fig:pdf_ei_eom} and \ref{fig:mean_ei_eom} thus show that
the alignment properties of $\mathbf{e}_\omega$ and $\mathbf{e}_2$, 
or the orthogonality 
between $\mathbf{e}_\omega$ and $\mathbf{e}_3$, are observable throughout 
the entire inertial range. 
This is to be contrasted with the lack of alignment between 
$\mathbf{e}_\omega$ and $\mathbf{e}_1$, which is observed over all the
scales that we have studied.

\subsection{The intermediate eigenvalues of the rate of strain}
\label{subsec:beta}

We now turn to the statistics of $\strain_i$, the eigenvalues of the rate of strain.
Whereas it is clear, from the relations $ \strain_1 \ge \strain_2 \ge \strain_3 $ and 
$ \strain_1 + \strain_2 + \strain_3 = 0$ that $\strain_1 \ge 0$ and $\strain_3 \le 0$, the sign of 
$\strain_2$ can be either positive or negative. To quantify the relative value of 
$\strain_2$ compared to $\strain_1$ and $\strain_3$, it is convenient to 
introduce the dimensionless variable $\beta$, defined by \cite{Ashurst87}:
\begin{equation} 
\beta \equiv \frac{\sqrt{6} ~ \strain_2}{\sqrt{\strain_1^2 + \strain_2^2 + \strain_3^2} }.
\label{def_beta}
\end{equation} 
This definition ensures that $ -1 \le \beta \le 1$.

%
%
\begin{figure}
\begin{center}
\subfigure[]{
	\includegraphics[width=0.6\textwidth,angle=0]{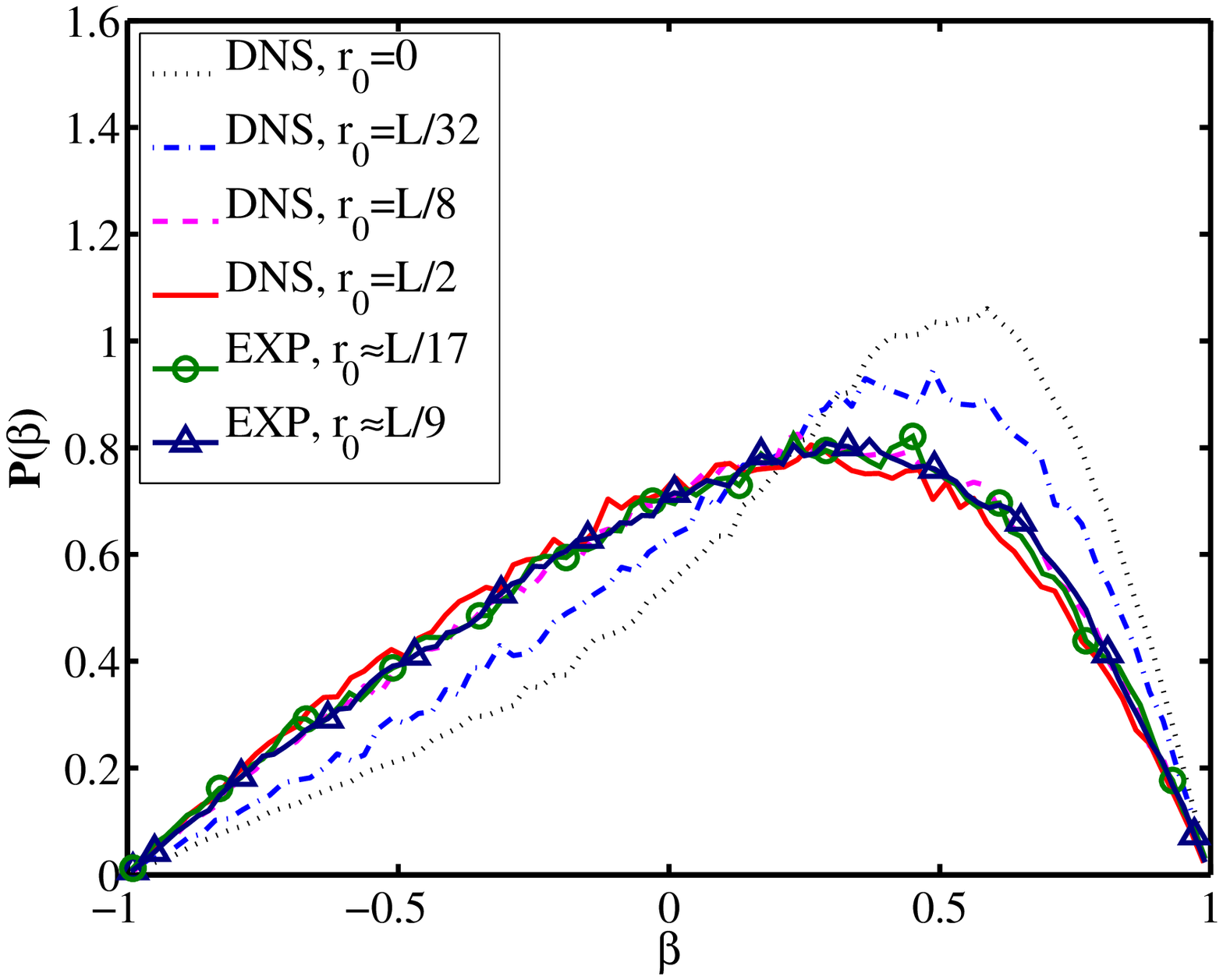}
}
\\
\subfigure[]{
	\includegraphics[width=0.6\textwidth,angle=0]{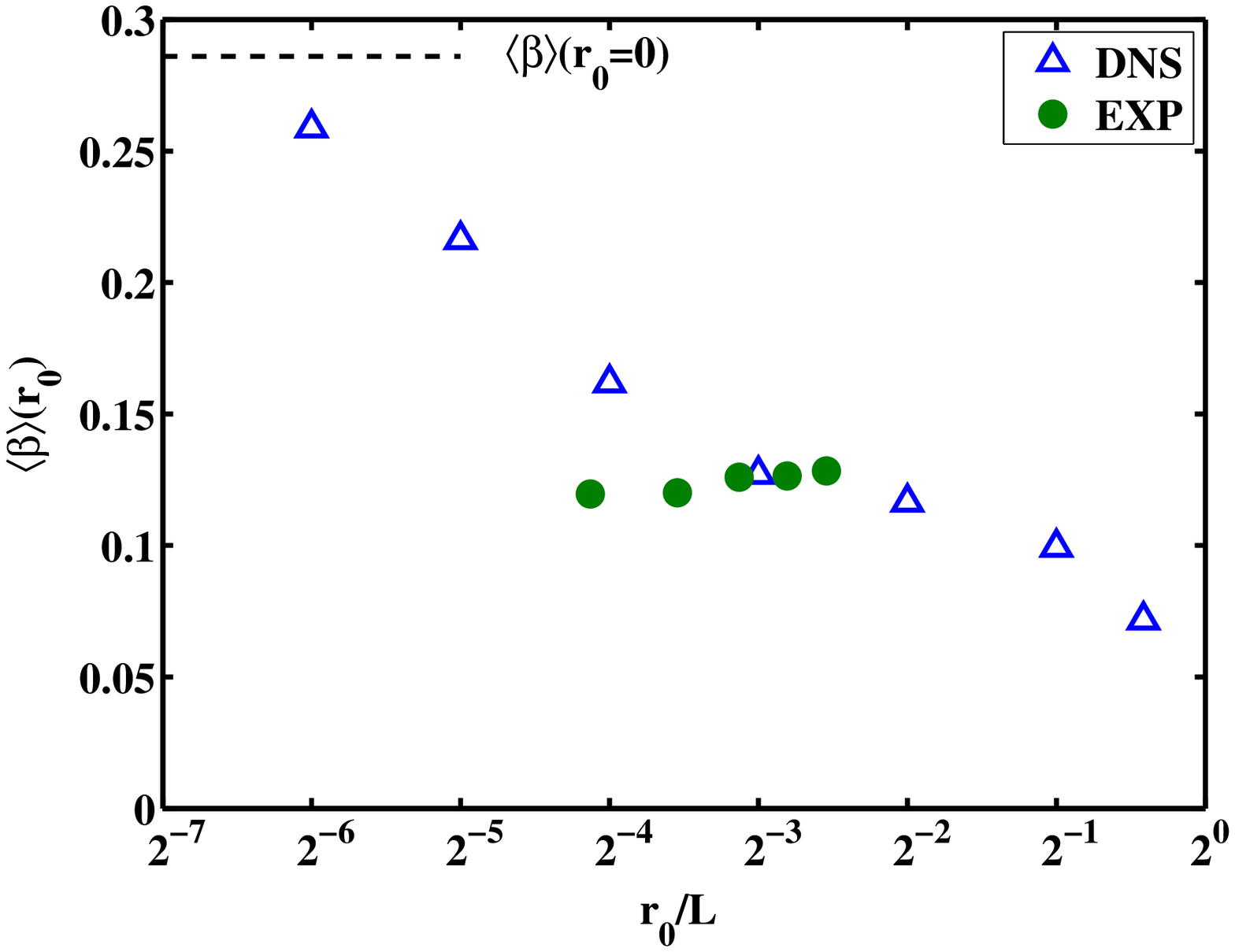}
}
\caption{(Color online) Dependence of the intermediate eigenvalue of strain on scale $r_0$.
(a) PDF of $\beta$, as defined in Eq.~\ref{def_beta}, for several values of $r_0$.
(b) The average value of $\beta$ as a function of scale, in which the dashed line indicates $\langle \beta \rangle$ for the true velocity gradient $\mathbf{m}$ as $r_0 \rightarrow 0$.
The DNS data correspond to $R_\lambda 
= 170$, whereas the experiments to $R_\lambda = 350$.
}
\label{fig:PDF_beta}
\end{center}
\end{figure}

The PDF of $\beta$ is shown in Fig.~\ref{fig:PDF_beta}a for several values of $r_0$. 
At small Reynolds number, the PDFs show a clear evolution as a function
of the scale $r_0$. In the dissipative
range ($r_0 = 0$, {\it i.e.}, for the true velocity gradient tensor), a 
strong bias towards high values of $\beta$ is observed.
As $r_0$ increases, the PDFs show a reduced bias towards high values of $\beta$.
For the highest Reynolds number experimental flow,  the PDF of $\beta$ 
changes little  throughout the limited range of measured scales.  
As shown in Fig.~\ref{fig:PDF_beta}b, the mean 
$\langle \beta \rangle$ remains positive throughout the entire
inertial range, dropping from $\langle \beta \rangle \approx 0.285$ 
for $ r_0 = 0$, to $\langle \beta \rangle \approx 0.07$ for 
$r_0 \approx L$ for the low Reynolds number data. On the contrary, 
the value for the higher Reynolds number experimental flow, in the range 
$1/17 \le r_0/L \le 1/5$ is close to $0.13$.
Figure~\ref{fig:PDF_beta} suggests that the dependence of 
$\langle \beta \rangle$ on scale is relatively weak in the
inertial range $L/16 \lesssim r_0 \lesssim L/2$, in particular
at the highest Reynolds number. This is consistent with the trend observed in
subsection \ref{subsec:algnmt_vort_strn}, which points to a 
scale-independent structure of $\mathbf{M}$ in the inertial range at sufficiently high Reynolds numbers.

The results of this section suggest a smooth change of the structure of $\mathbf{M}$ throughout the range of scales. 
Whereas no particular alignment between $\mathbf{e}_\omega$ and $\mathbf{e}_1$ is observed, irrespective
of the size of the tetrahedron, the tendency of $\mathbf{e}_\omega$ to align with $\mathbf{e}_2$ 
(respectively to be perpendicular to $\mathbf{e}_3$) is the strongest for $r_0 \rightarrow 0$, {\it i.e.}, 
for the true velocity gradient tensor $\mathbf{m}$, and decreases for $\mathbf{M}$ with
increasing $r_0$. 
Similarly, the intermediate eigenvalue of the rate of strain is comparatively more positive for 
the true velocity gradient tensor $\mathbf{m}$ than for $\mathbf{M}$ at values of $r_0$ in the inertial range.
As the scale $r_0 $ increases towards the inertial length $L$, the PDF of $\beta$ becomes more symmetric about $\beta = 0$, as expected from a random matrix.

Thus, the perceived velocity gradient tensor $\mathbf{M}$ in the inertial range shares many essential properties with the true velocity gradient tensor $\mathbf{m}$.
Our observations thus suggest a  continuity in the behavior from the dissipative 
to the inertial range. Remarkably,
the data obtained with the higher Reynolds number experimental flow suggest that the structural
properties of $\mathbf{M}$ become scale independent in the inertial scale. This
should be confirmed by studying flows at even higher Reynolds numbers.

{
\section{Properties of M: Dynamics (Temporal correlations)}
\label{sec:dynamics}
}

The main observation in our recent study \cite{XPB11} was the dynamical alignment of vorticity with the leading eigenvector of the rate of strain, which we were able to study  conveniently by investigating the correlation between $\mathbf{e}_1(0)$ and $\mathbf{e}_\omega(t)$, \ie, with a time delay.
Hence the lack of alignment between $\mathbf{e}_\omega(t)$ and $\mathbf{e}_1(t)$ (Figs.~\ref{fig:pdf_ei_eom}(a) and (b))
is a consequence of the de-correlation of the vector $\mathbf{e}_1$ 
itself in a time $t$, which {counteracts} the alignment of 
$\mathbf{e}_\omega(t)$ with $\mathbf{e}_1(0)$. 

A consequence of the previous section is that it is not \textit{a priori} clear that the
alignment of $\mathbf{e}_\omega(t)$
with {$\mathbf{e}_i(0)$ ($i = 2$ or $i = 3$)} is going to be self-similar,
\ie, that the angle between $\mathbf{e}_\omega(t)$ and
{$\mathbf{e}_i(0)$ ($i = 2$ or $i = 3$)} is going to depend only on $t/t_0$, as it had
been found for the angle between $\mathbf{e}_\omega(t)$ and
$\mathbf{e}_1(0)$ \cite{XPB11}. 
This is due to the fact that the PDFs of $| \mathbf{e}_\omega(0) \cdot 
\mathbf{e}_i (0)|$, for $i = 2 $ and $i=3$ change with the scale $r_0$ for small scale separation $L/\eta$.

\subsection{Averaged alignment of the direction of vorticity with the eigenvectors of the rate of strain}
\label{subsec:averaged_alignment}

%
%
\begin{figure}
\begin{center}
\subfigure[]{
	\includegraphics[width=0.49\textwidth,angle=0]{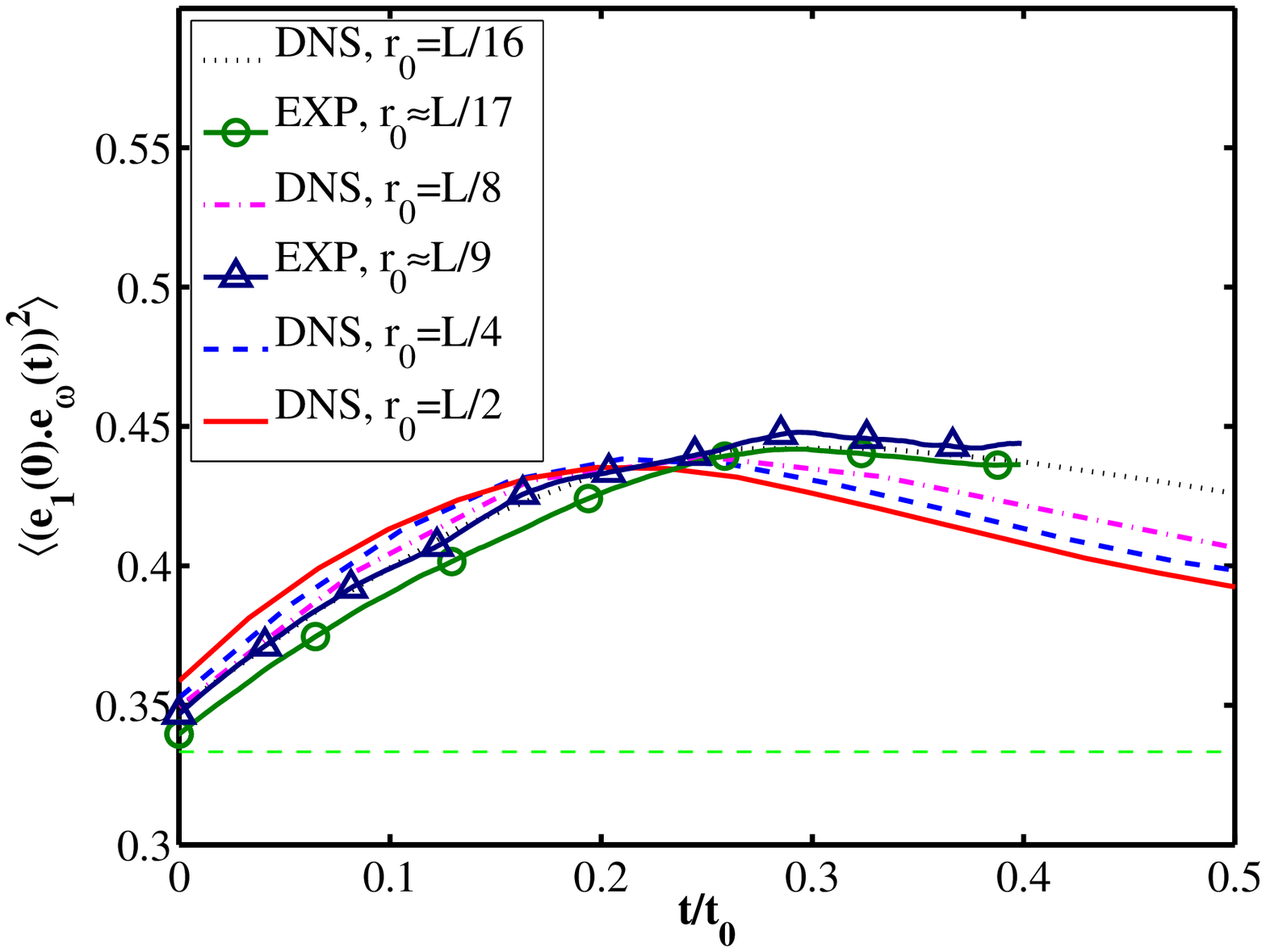}
}
\\
\subfigure[]{
	\includegraphics[width=0.49\textwidth,angle=0]{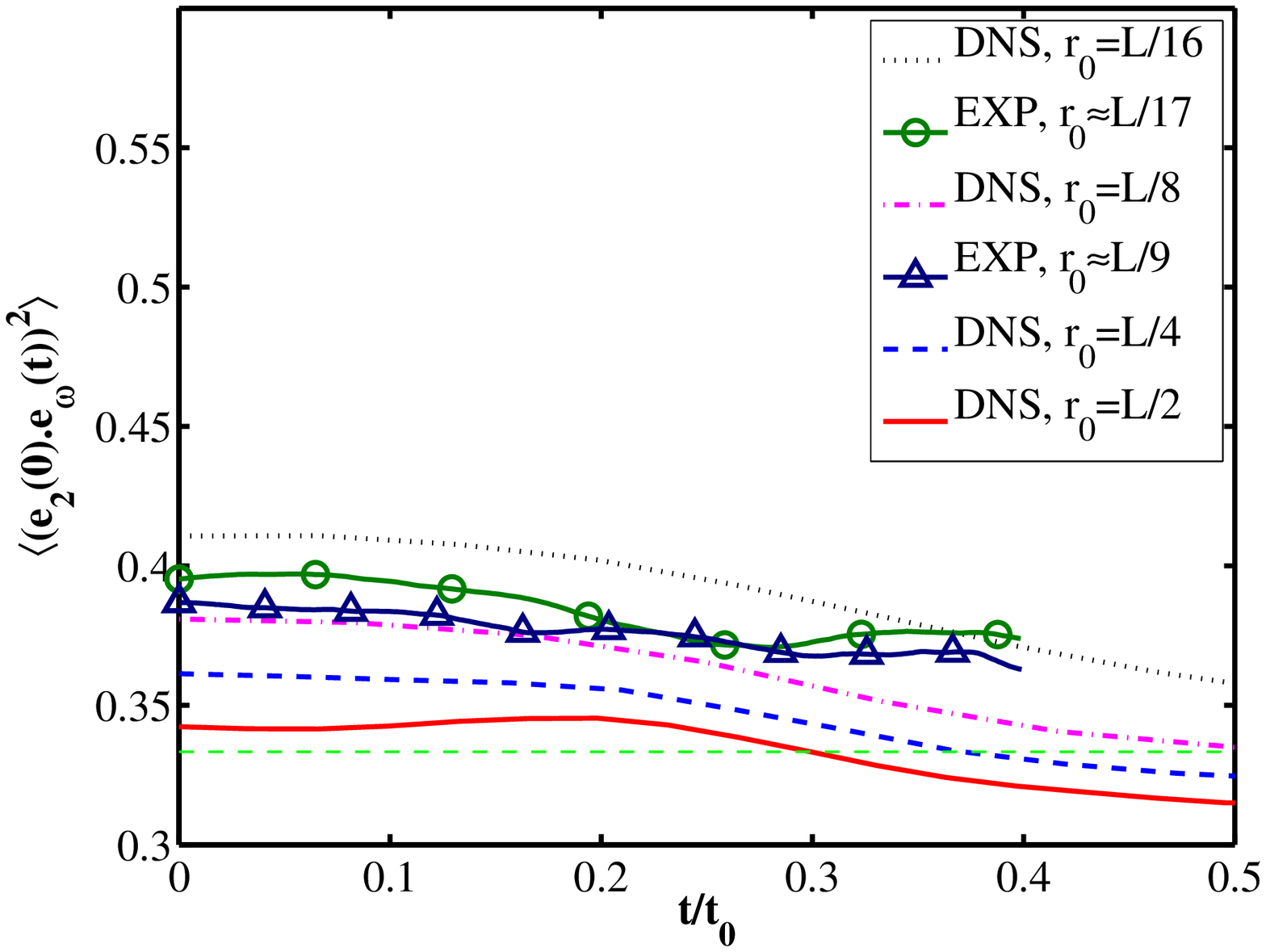} 
}
\\
\subfigure[]{
	\includegraphics[width=0.49\textwidth,angle=0]{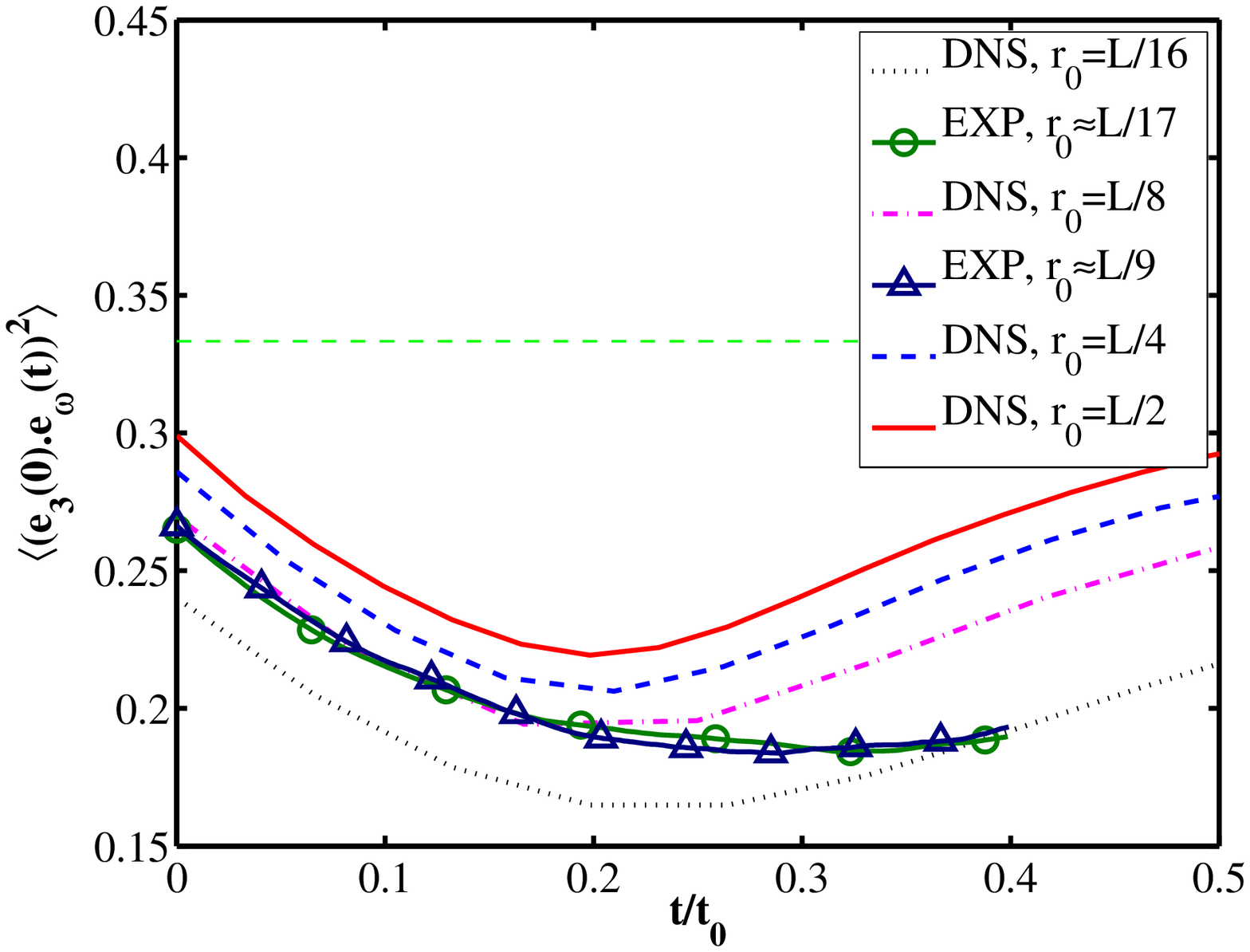}
}
\caption{(Color online) Alignment between $\mathbf{e}_\omega(t)$ and $\mathbf{e}_i(0)$. 
Ensemble averages $\langle [\mathbf{e}_i(0) \cdot \mathbf{e}_\omega(t) ]^2 \rangle$
for $i=1$ (a), $i=2$ (b) and $i=3$ (c) as a function of the time delay $t$, normalized by $t_0$. 
The initial tetrahedron size, $r_0$, are all in the inertial range.
The evolution of the alignment of $\mathbf{e}_\omega(t)$ with 
$\mathbf{e}_1(0)$ suggests a self-similarity: The curves in (a) all superpose once $t$ is expressed in units of $t_0$.
The alignment of $\mathbf{e}_\omega(t)$ with $\mathbf{e}_2(0)$ and $\mathbf{e}_3(0)$ shows a stronger
dependence on $r_0$, consistent with the scale dependence of 
the {{\it instantaneous}} alignment properties of $\mathbf{e}_\omega$ with $\mathbf{e}_2$ and $\mathbf{e}_3$ (Fig.~\ref{fig:pdf_ei_eom}).
The DNS data correspond to $R_\lambda 
= 170$, whereas the experiments to $R_\lambda = 350$.
}
\label{fig:dyn_align_eo_ei}
\end{center}
\end{figure}

%
%
\begin{figure}
\begin{center}
\subfigure[]{
	\includegraphics[width=0.31\textwidth,angle=0]{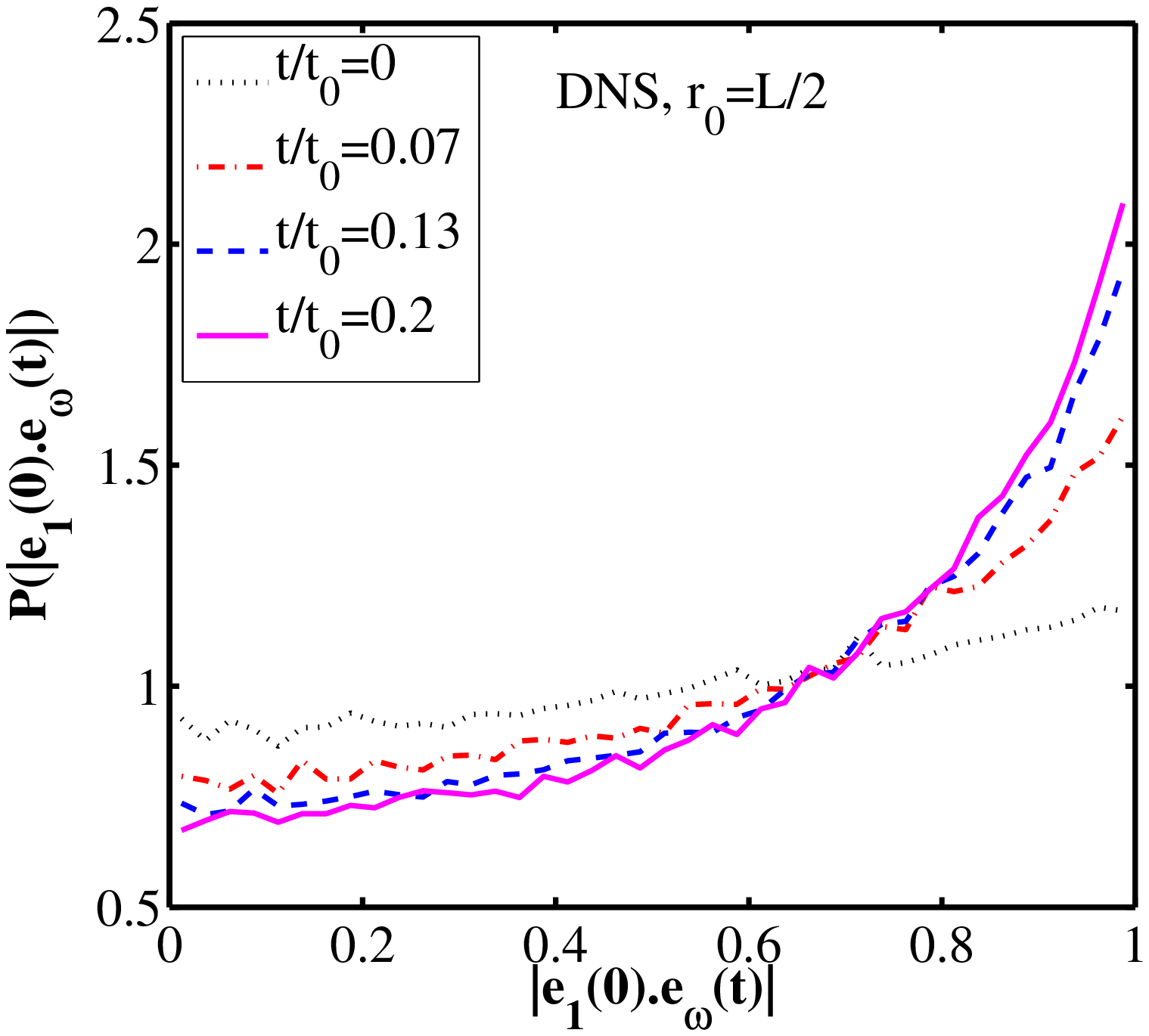}
}
\subfigure[]{
	\includegraphics[width=0.31\textwidth,angle=0]{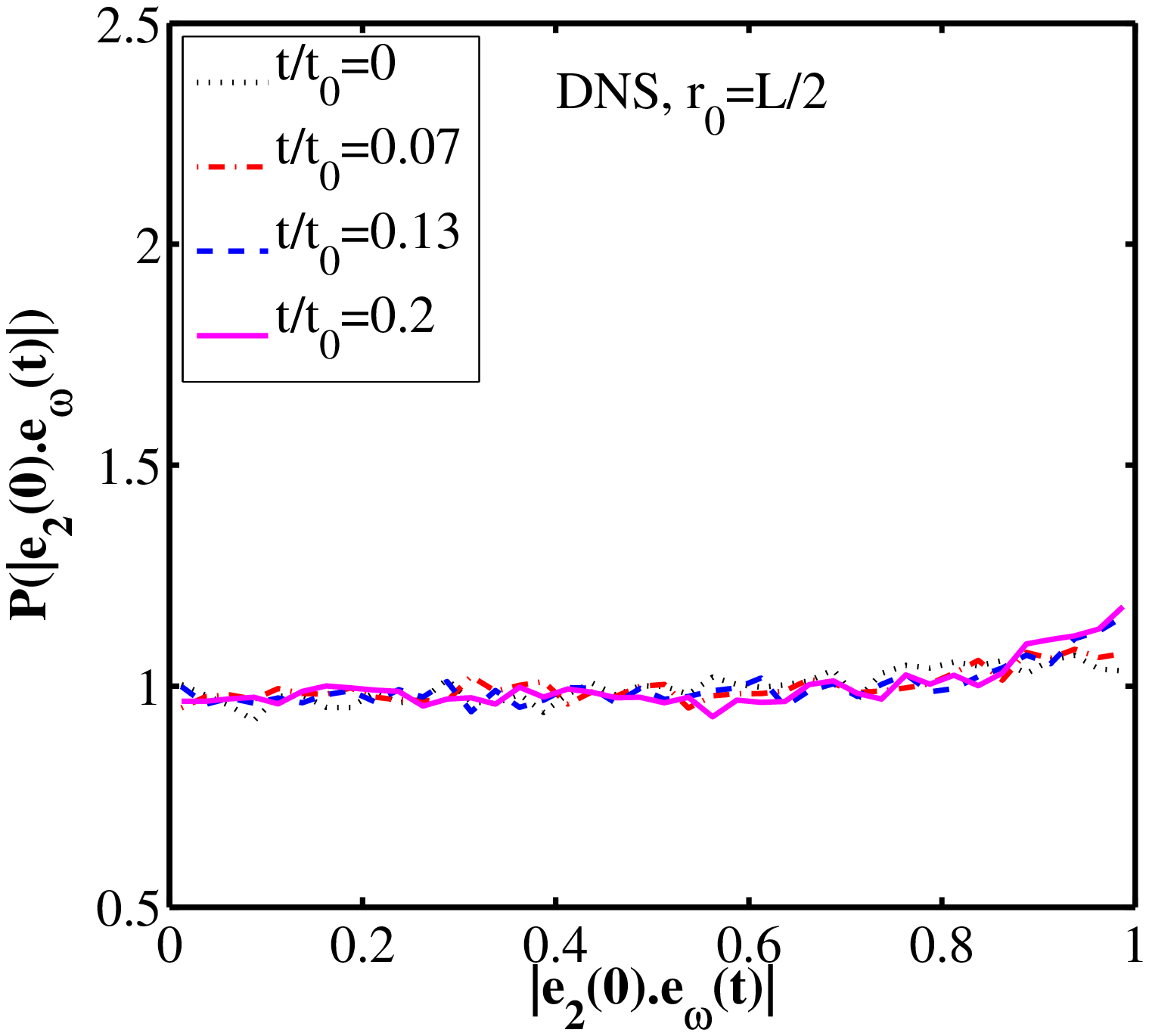}
}
\subfigure[]{
	\includegraphics[width=0.31\textwidth,angle=0]{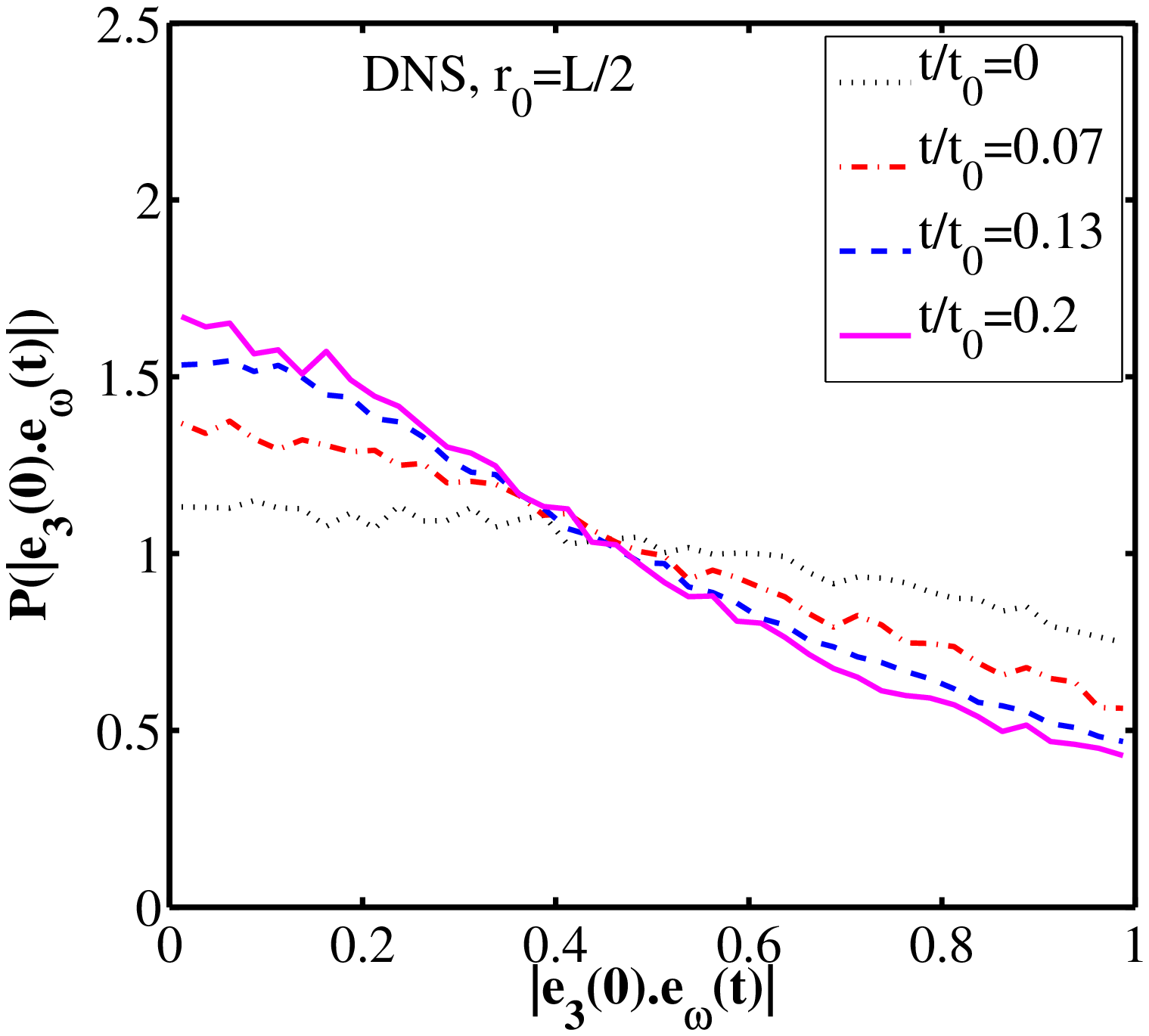}
}
\\
\subfigure[]{
	\includegraphics[width=0.31\textwidth,angle=0]{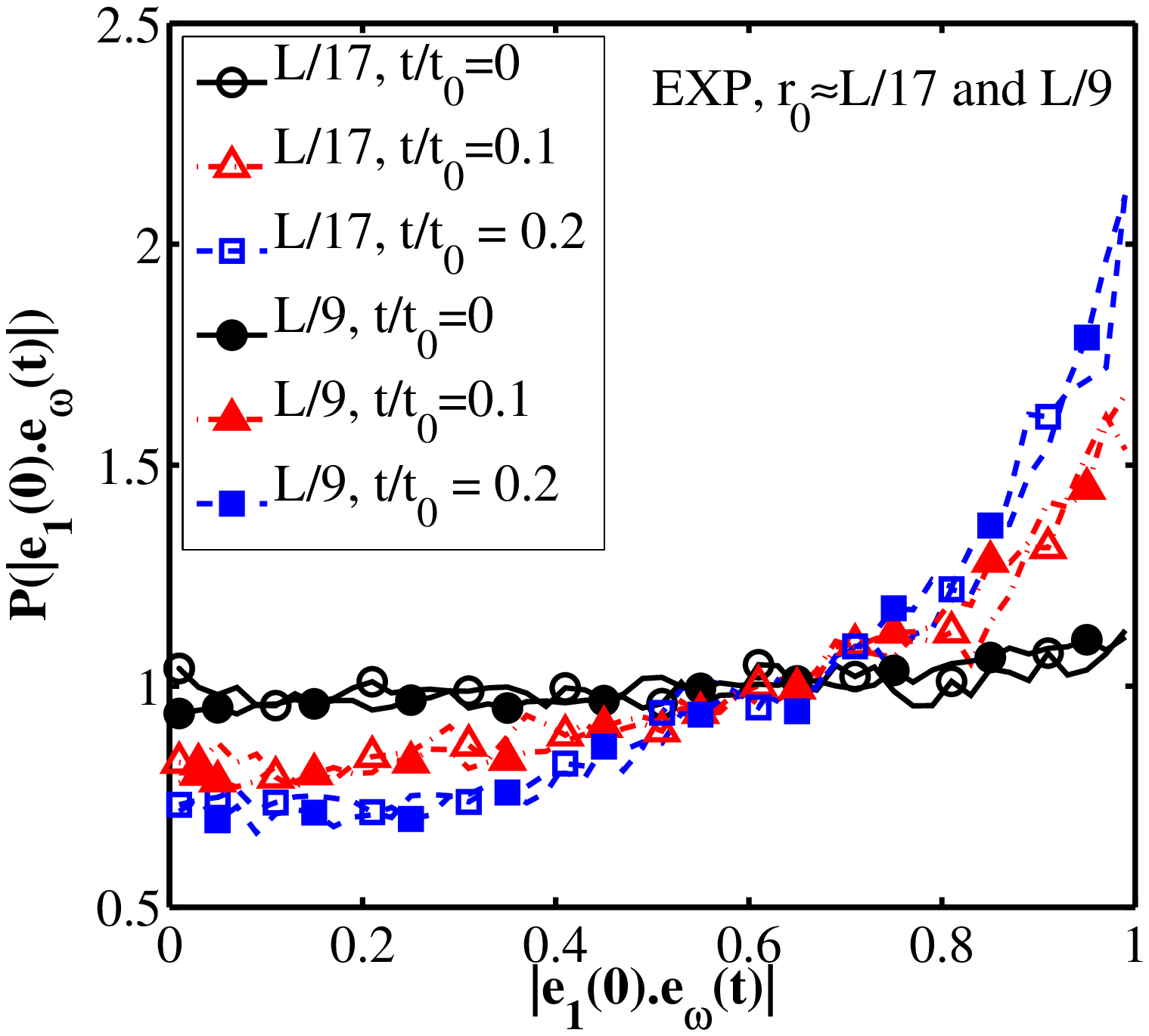}
}
\subfigure[]{
	\includegraphics[width=0.31\textwidth,angle=0]{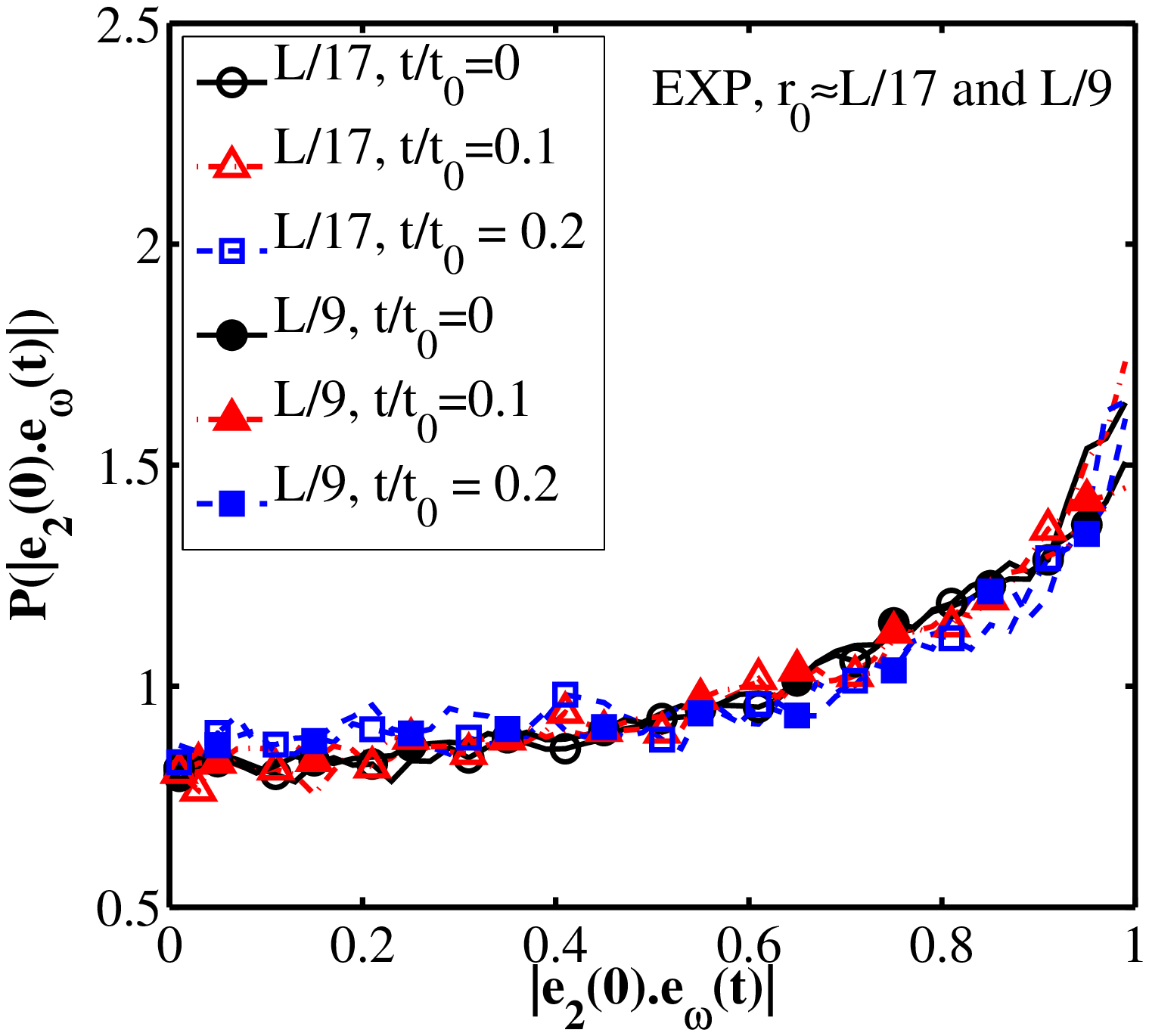}
}
\subfigure[]{
	\includegraphics[width=0.31\textwidth,angle=0]{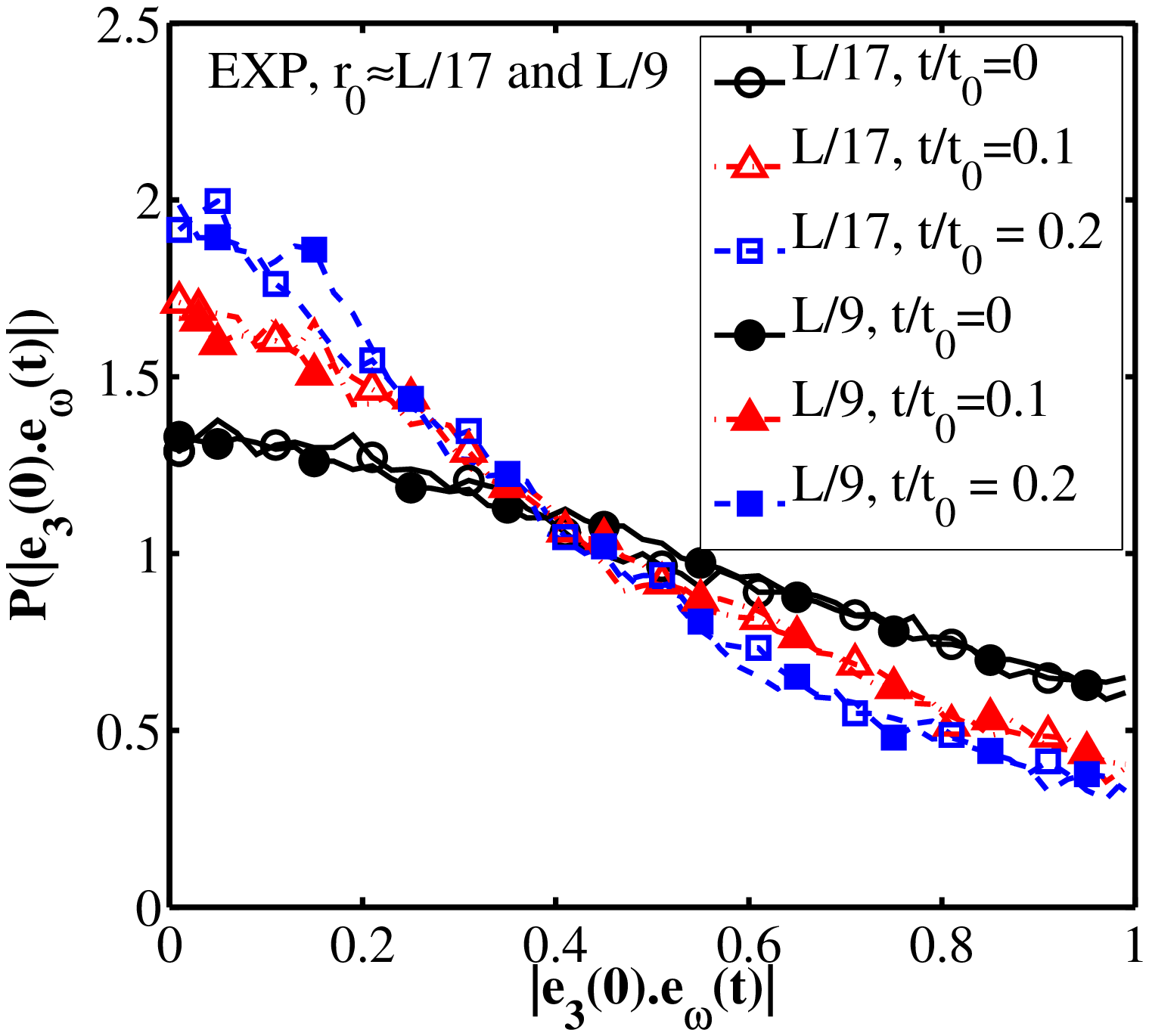}
}
\\
\subfigure[]{
	\includegraphics[width=0.31\textwidth,angle=0]{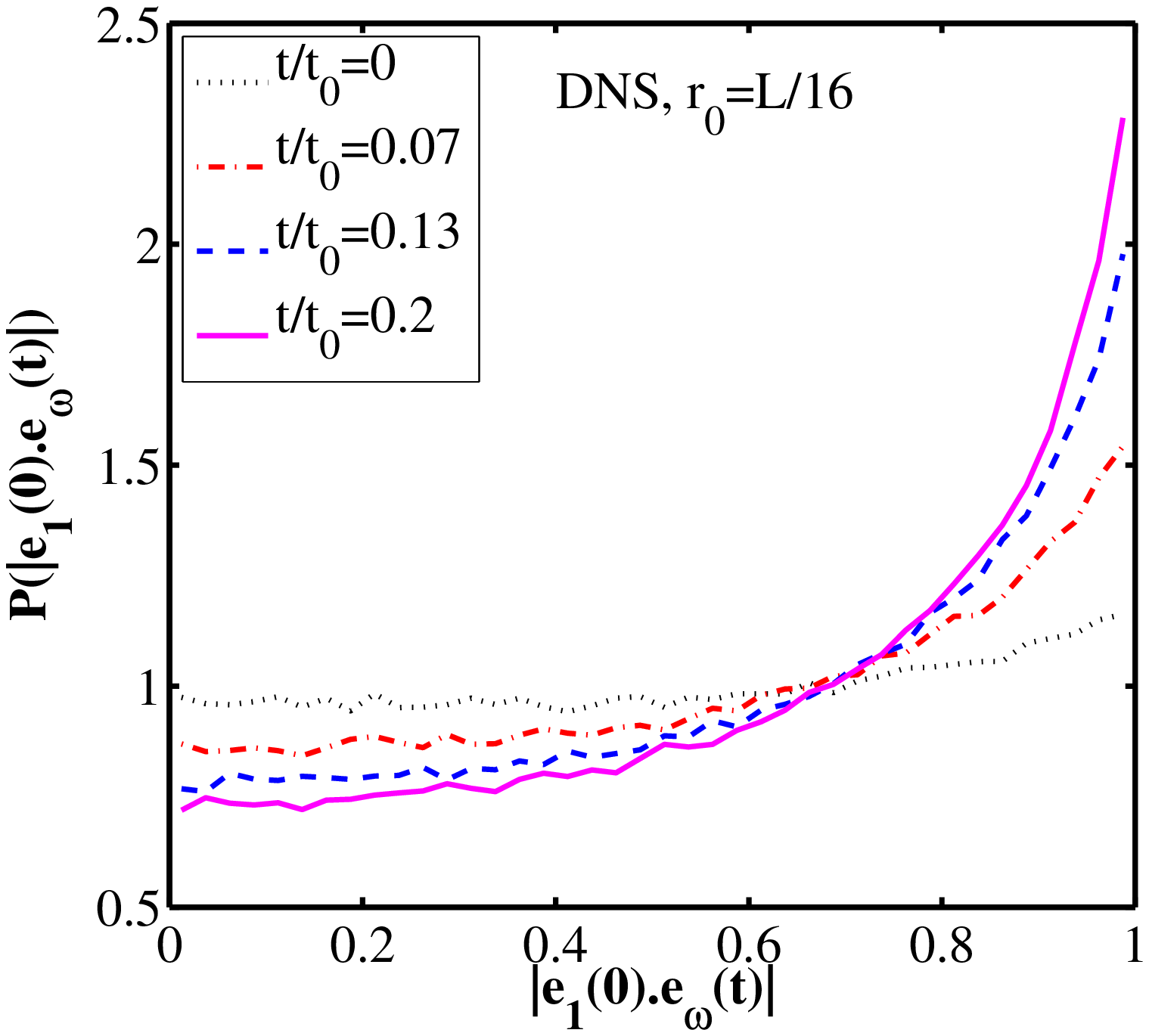}
}
\subfigure[]{
	\includegraphics[width=0.31\textwidth,angle=0]{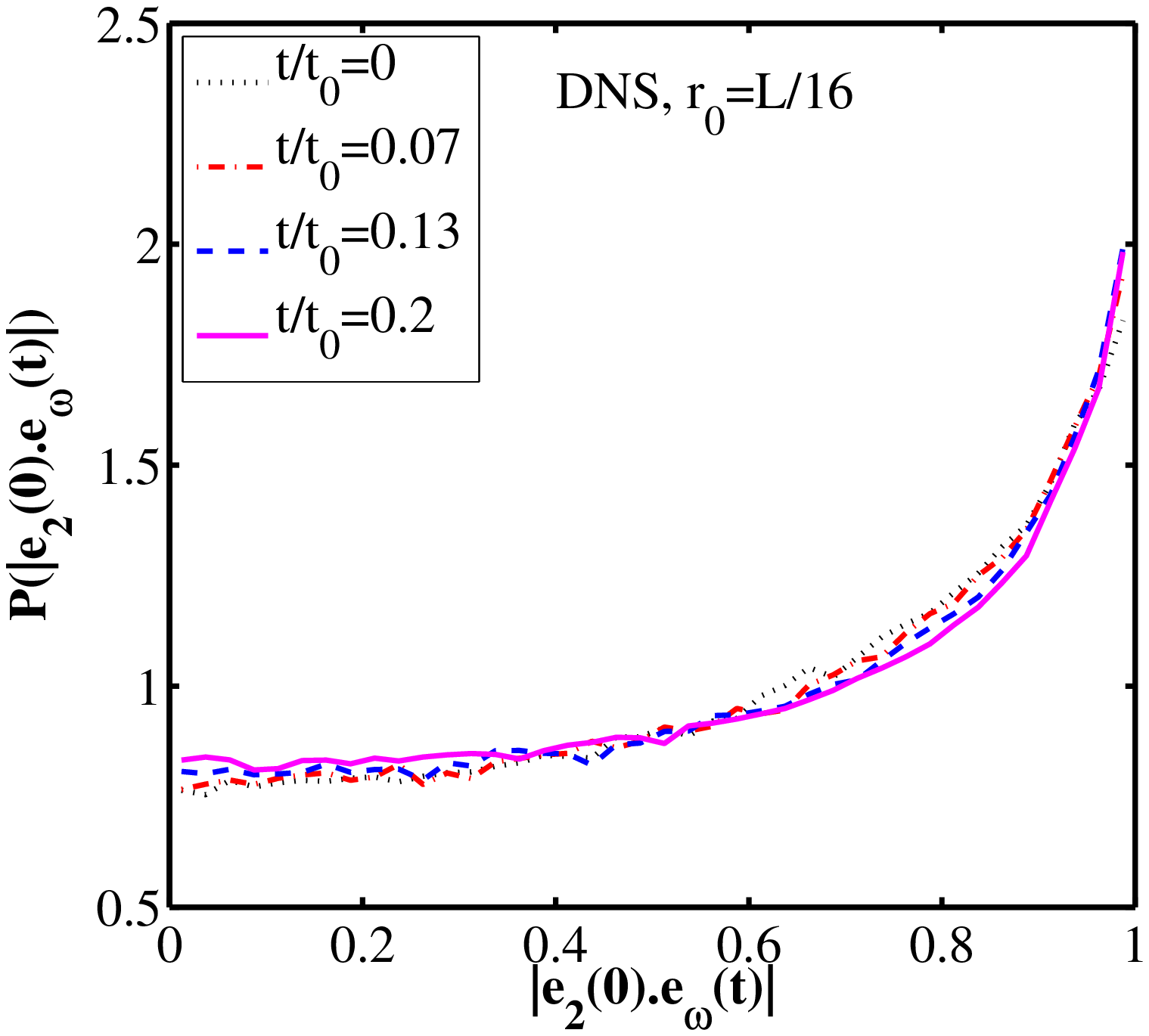}
}
\subfigure[]{
	\includegraphics[width=0.31\textwidth,angle=0]{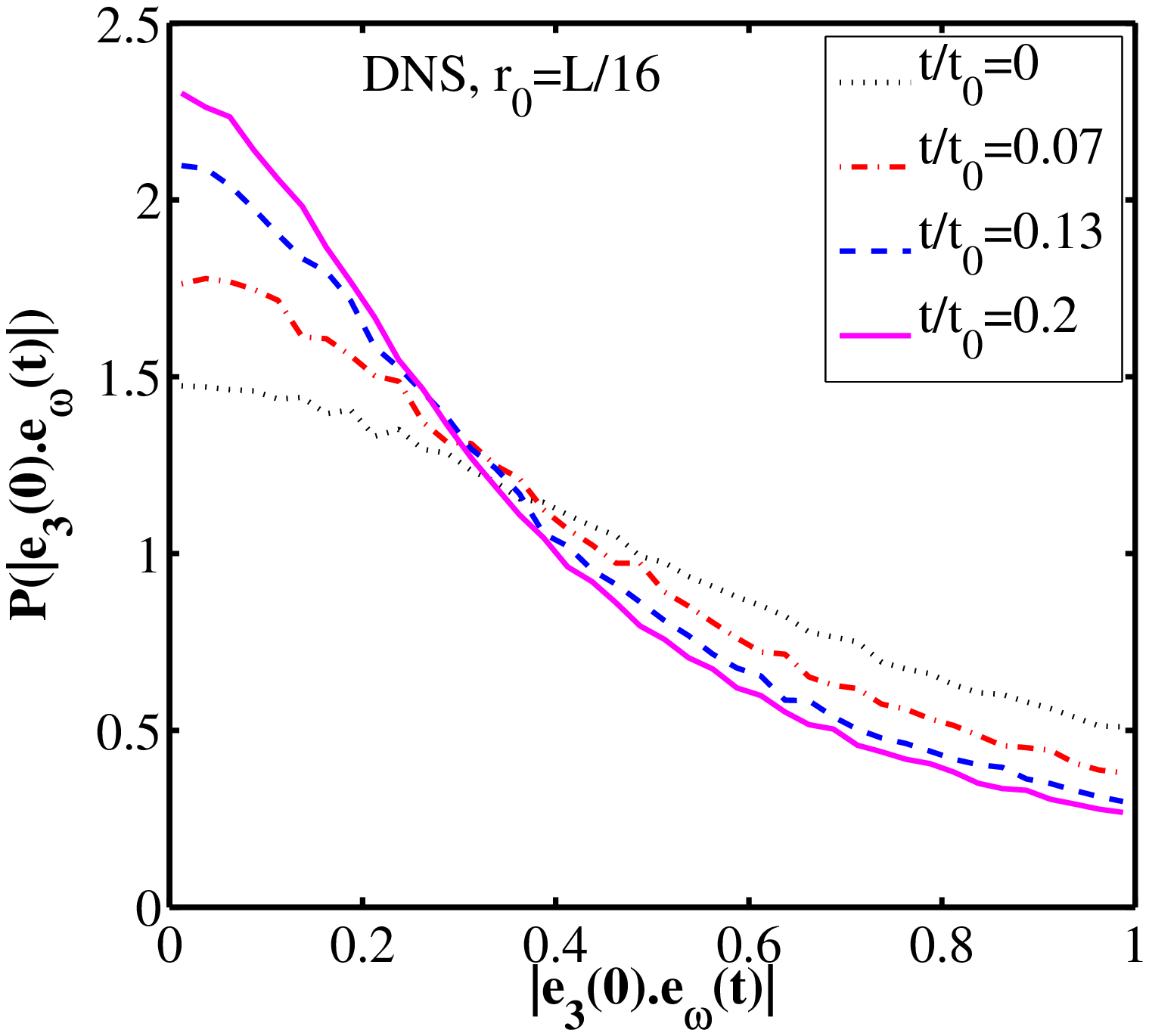}
}
\caption{(Color online) Evolution of the angle between $\mathbf{e}_i(0)$ and $\mathbf{e}_\omega(t)$.
The PDF of $| \mathbf{e}_i(0) \cdot \mathbf{e}_\omega(t)| $ is shown for times between $0 \leq t/t_0 \leq 0.2$, 
and at three values of $r_0$: $r_0 = L/2$ in DNS ($R_\lambda = 170$) (a)-(c),  $r_0 \approx L/9$ and $r_0 \approx L/17$ in experiment ($R_\lambda = 350$) (d)-(f), and $r_0 = L/16$ in DNS (g)-(i). 
Panels (a), (d) and (g) are for $i=1$. Consistent with previous results \cite{XPB11}, the evolution of these PDFs is essentially self-similar for $r_0$ in the inertial range. 
Panels (b), (e) and (h) are for $i=2$. The PDFs hardly change over the range of times shown; however, they depend significantly on the value of $r_0$.
Panels (c), (f) and (i) are for $i=3$. During the period of time shown, the PDFs evolve to peak at $0$, {\it i.e.}, 
$\mathbf{e}_\omega(t)$ tends to become perpendicular to $\mathbf{e}_3(0)$. There is also a moderate dependence on scale $r_0$.
}
\label{fig:angle_ei_eom}
\end{center}
\end{figure}

Fig.~\ref{fig:dyn_align_eo_ei}(a) shows {that, as time increases, the
vector $\mathbf{e}_\omega (t)$ tends to become better aligned with
$\mathbf{e}_1(0)$.  
As reported previously~\cite{XPB11}, the statistics characterizing the
alignment of the two vectors is essentially independent of the
scale $r_0$, once time $t$ is rescaled by $t_0$. The characteristic
time scale for this process, 
Fig.~\ref{fig:dyn_align_eo_ei}(a) is of order $t_0/5$.}
For the DNS at $R_\lambda = 170$, the alignment of $\mathbf{e}_\omega(t) $ with $\mathbf{e}_2(0)$, 
and to a lesser extend with $\mathbf{e}_3(0)$, shows a significant dependence on scale,
see Fig.~\ref{fig:dyn_align_eo_ei}(b) and (c).
Note that for $t=0$, the difference was shown as a function of $r_0/L$ in Fig.~\ref{fig:mean_ei_eom}. 
In comparison, the experimental data obtained at $R_\lambda = 350$
shows a much reduced variation as a function of scale (see also Fig.~\ref{fig:mean_ei_eom}).
We also notice that $\langle [ \mathbf{e}_2(0) \cdot \mathbf{e}_\omega(t) ]^2 \rangle$
is almost constant, \ie, the slope at $t = 0$ is much smaller than the
slopes for $\mathbf{e}_1(0)$ and $\mathbf{e}_3(0)$. These features can
be understood by the elementary considerations
presented in subsection
\ref{subsection:modeling}.

{We note that the time scale of the alignment of $\mathbf{e}_\omega(t)$ with
$\mathbf{e}_1(0)$ and $\mathbf{e}_3(0)$ are comparable to the
time scale characterizing the deformation of the tetrahedra 
(see Section~\ref{sec:deformation}). 
This is a strong indication that the deformation is an important part of the 
dynamical process, see also~\cite{tennekes:1972,XPB11}. A theorical description
of the dynamics of $\mathbf{M}$ has therefore to take into account in
an essential way the deformation of the tetrahedra~\cite{CPS99}. }

Not only the mean values of $[ \mathbf{e}_1(0) \cdot \mathbf{e}_\omega(t) ]^2 $ 
{evolve as a function} of $t/t_0$ in a self-similar way, but also the PDFs of $\vert \mathbf{e}_1(0) \cdot \mathbf{e}_\omega(t) \vert $ (see Figures~\ref{fig:angle_ei_eom}(a), (d), and (g)). 
This situation is to be contrasted with the evolution of the alignment between $\mathbf{e}_\omega(t)$ and $\mathbf{e}_2(0)$ 
or $\mathbf{e}_3(0)$.
Figures~\ref{fig:angle_ei_eom}(b), (e), and (h) show that the PDF of the angle 
between $\mathbf{e}_2(0)$ and $\mathbf{e}_\omega(t)$ is essentially independent of $t/t_0$ for $0 \le t/t_0 \le 0.2$, consistent with Fig.~\ref{fig:dyn_align_eo_ei}(b). 
The direction $\mathbf{e}_\omega(t)$ is strongly aligned with $\mathbf{e}_2(0)$
at small values of $r_0$, much less so as the value of $r_0$ increases towards $L$. 
Last, as shown in Figs.~\ref{fig:angle_ei_eom}(c), (f), and (i), 
$\mathbf{e}_\omega (t)$ becomes more perpendicular to $\mathbf{e}_3(0)$ within $t/t_0 \lessapprox 0.2$ and
the tendency is stronger at small values of $r_0$. 

In summary, the main observation of this subsection is that, contrary to the alignment
of $\mathbf{e}_\omega(t)$ with $\mathbf{e}_1(0)$, which evolves in an essentially 
self-similar manner, the dynamics of the alignment of $\mathbf{e}_\omega(t)$ with $\mathbf{e}_2(0)$ and $\mathbf{e}_3(0)$
proceeds in a way that depends on the size of the tetrahedra, at least for our Reynolds numbers,
a result that was expected
in view of the observation that the average values of 
$( \mathbf{e}_i(0) \cdot \mathbf{e}_\omega(0))^2 $ show a clear dependence on scale.

\subsection{Elementary modeling considerations}
\label{subsection:modeling}

The analysis presented below, which aims only at capturing some of the main features of the alignment process, rests on several important assumptions. We restrict ourselves here to very short time scales, which effectively allows us to neglect the coupling between shape and vorticity. We are merely investigating here the evolution of vorticity, at the lowest order in terms of a Taylor-series expansion. Effectively, coupling terms between shape and vorticity, as well as feedback between vorticity and strain all would be seen only at higher order, independently of the form of the model. The limitation of this approach is that it does not allow us to properly describe the coupling between shape deformation and the evolution of vorticity.
To describe the evolultion of $\mathbf{M}$, models based
on the Restricted Euler approximation~\cite{Vieill82,Vieill84,Cantwell92} 
and its subsequent elaborations, reviewed in~\cite{Meneveau11}, have been 
used very often. The resulting equation for the evolution of the vorticity is:
\begin{equation}
\frac{d \omega_i}{dt} - S_{ij} \omega_j = 0.
\label{eq:domegadt_REuler}
\end{equation}
which can also be obtained directly from the Euler equation.
A full description of $\mathbf{M}$ requires
also an equation for the evolution of the rate of strain, $\mathbf{S}$,
which 
involves the second derivatives of pressure (pressure
hessian), a quantity notoriously difficult to model. While it may be possible to
extract information concerning the strain evolution from the available data,
we restrict ourselves here to the evolution of vorticity, and focus
exclusively on Eq.~\ref{eq:domegadt_REuler}.  

An evolution equation for the direction of vorticity $\mathbf{e}_\omega$ can
be readily obtained by taking the double vector product of Eq.~\ref{eq:domegadt_REuler} with $\mathbf{e}_\omega$, which immediately leads to:
\begin{equation}
\frac{d \mathbf{e}_\omega}{dt} = \mathbf{S} \cdot \mathbf{e}_\omega - 
( \mathbf{e}_\omega \cdot [\mathbf{S} \cdot \mathbf{e}_\omega] ) \mathbf{e}_\omega
\label{eq:d_eomega_dt_REuler}
\end{equation}
To proceed, we project $\mathbf{e}_\omega$ on the three eigenvectors
of $\mathbf{S}$: 
\begin{equation}
\mathbf{e}_\omega = \sum_{i=1}^3 c_i \times \mathbf{e}_i
\label{eq:project}
\end{equation}
The constraint $|\mathbf{e}_\omega|^2 = 1$ implies that 
$\sum_{i=1}^3 c_i^2 = 1$.
Eq.~\ref{eq:d_eomega_dt_REuler} can be simply written in terms of the three
components $c_i$. As we are  primarily interested here in 
$\langle (\mathbf{e}_\omega \cdot \mathbf{e}_i )^2 \rangle$, 
we use Eq.~\ref{eq:d_eomega_dt_REuler} for $c_i^2$:
\begin{equation}
\frac{d c_i^2}{dt} \bigg \vert_{t=0}= 2 c_i^2 \bigg[ \strain_i - \sum_{j=1}^3 \strain_j c_j^2 \bigg].
\label{eq:ci_sq_diff}
\end{equation}

The initial dynamics of alignment is best investigated by averaging over many
realizations, and by separating the three components in Eq. \eqref{eq:ci_sq_diff}:
\begin{equation}
\bigg \langle \frac{d c_1^2}{dt} \bigg \rangle \bigg \vert_{t=0} = 
2 \langle c_1^2 c_2^2(\strain_1 - \strain_2) \rangle 
+ 2 \langle c_1^2 c_3^2(\strain_1 - \strain_3) \rangle
\label{eq:dc1_dt}
\end{equation}
\begin{equation}
\bigg \langle \frac{d c_2^2}{dt} \bigg \rangle \bigg \vert_{t=0} = 
- 2 \langle c_1^2 c_2^2(\strain_1 - \strain_2) \rangle 
+ 2 \langle c_3^2 c_2^2 (\strain_2 - \strain_3) \rangle
\label{eq:dc2_dt}
\end{equation}
and
\begin{equation}
\bigg \langle \frac{d c_3^2}{dt} \bigg \rangle \bigg \vert_{t=0} = 
-2 \langle c_1^2 c_3^2(\strain_1 - \strain_3) \rangle
- 2 \langle c_2^2 c_3^2 (\strain_2 - \strain_3)  \rangle
\label{eq:dc3_dt}
\end{equation}
The sign of $\langle\frac{d c_i^2}{dt}\rangle \vert_{t=0}$ is determined solely by the eigenvalues of the rate of strain, with  
$\strain_1 \geq \strain_2 \geq \strain_3$. We see immediately that  $\langle\frac{d c_1^2}{dt}\rangle \vert_{t=0}>0$ and 
$\langle\frac{d c_3^2}{dt}\rangle \vert_{t=0}<0$. Thus initially vorticity tends to align with $\mathbf{e}_1(0)$ and to become perpendicular to $\mathbf{e}_3(0)$. By the following argument we also see that  $\langle\frac{d c_2^2}{dt}\rangle \vert_{t=0}$ is smaller in magnitude than the other two components.  The first term on the RHS of  Eq.~\ref{eq:dc2_dt} is negative and has the same magnitude as  the first term on the RHS of Eq.~\ref{eq:dc1_dt}, hence this term is negative and smaller  in magnitude than $\langle\frac{d c_1^2}{dt}\rangle \vert_{t=0}$. Also, the second term on the RHS of Eq.~\ref{eq:dc2_dt} is positive and has the same magnitude as the second term in the RHS of Eq.~\ref{eq:dc3_dt},   therefore this term is smaller than the magnitude  of $\langle\frac{d c_3^2}{dt}\rangle \vert_{t=0}$.  
Thus $\langle\frac{d c_2^2}{dt}\rangle \vert_{t=0}$ is smaller in magnitude 
than the other two terms and consequently its evolution is the slowest. 
This agrees qualitatively with the observations reported before.

The analysis presented above is merely based on the action of strain
on the vorticity, which is common to essentially all models aimed at describing
the evolution of $\mathbf{M}$. Comparing more quantitatively the results of 
the present work with the prediction at short times of more sophisticated models, such as 
the tetrahedron model~\cite{CPS99} or the Lagrangian stochastic 
model~\cite{Chevillard06,ChevMen11}, requires the determination of
several constants in the models, which is beyond the scope of this work.  
We nevertheless note that in the specific case of the tetrahedron model \cite{CPS99}, the equation of evolution of vorticity reads:
\begin{equation}
\frac{ d \omega_i}{dt} - ( 1 - \alpha) S_{ij} \omega_j = 0
\label{eq:vort_CPS}
\end{equation}
where $\alpha$, which describes the reduction of nonlinearity, is a parameter that a-priori depends on the scale $r_0$. Our observation that the evolution of $\mathbf{M}$ may be self-similar at large Reynolds numbers throughout the inertial range suggests that the parameter of the model are independent of the scale $r_0$.

As explained already,
an extension of this calculation to later times requires both 
the knowledge of the interplay between vorticity
and the rate of strain\cite{She91,dresselhaus:1991,nomura:1998},
as well as the
geometry of the tetrahedra.
The very good superposition of the correlations 
$ \langle ( {\mathbf e}_\omega (t) \cdot {\mathbf e}_1 (0) )^2 \rangle $
in Fig.~\ref{fig:dyn_align_eo_ei} corresponding to tetrahedra of different sizes, while $r_0$ is in the inertial range,
is consistent with a description of the dynamics of $\mathbf{M}$ by such a model
based on the Restricted Euler approximation or one of its generalizations\cite{Vieill82,Cantwell92,Chevillard06,Meneveau11}. Indeed, on general grounds, $\mathbf{M}$
is expected to behave with scale as $M \propto (\varepsilon/r_0^2)^{1/3}$, 
{so the eigenvalues of the rate of strain $\lambda_i$ scale as $\lambda_i \propto
(\varepsilon /r_0^2)^{1/3}$, which in turn is consistent that the dynamics
of alignment proceeds with the characteristic time scale $t_0$,
as observed before \cite{XPB11} and in this work. More generally,
a dynamical equation of the form $\frac{d \mathbf{M}}{dt} \propto \mathbf{M}^2$ immediately
suggests that the characteristic time scale of the dynamics 
should be $t_0$, thus suggesting that all the dynamical process should depend
on $t/t_0$, at least when $r_0$ is in the inertial range. }
While our results on the alignment between ${\mathbf e}_\omega(t)$ and
${\mathbf e}_1(0)$ strongly suggests some universal behavior
as a function of scale $r_0$, it is not settled by the present work whether the dynamics of alignment 
of vorticity with the other two eigendirections of strain is also self-similar 
over a range of scales in the inertial scales. 
It would be very interesting to determine whether a dependence on 
scale $r_0$ of the dynamics of $\mathbf{M}$ persists, even at very large Reynolds numbers. 
These issues are important in many aspects for the modeling
of the flow at scales $r_0$, in particular concerning the applicability of
models based on Restricted Euler, and its generalizations.

\subsection{Alignment of vorticity with the {eigenvectors} of the rate of strain: conditional statistics}
\label{subsec:conditioned_averaged_alignment}

The instantaneous alignment of vorticity with the intermediate eigenvalue of the rate of strain,
documented many times before for the true velocity gradient tensor $\mathbf{m}$, and for the perceived velocity gradient $\mathbf{M}$
(see Fig.~\ref{fig:pdf_ei_eom}), results from an averaging over all 
configurations in the flow, which gives the same statistical weight 
to any configuration. This may lead to the impression that, in the case
of the velocity gradient tensor, the direction of the intermediate eigenvalue of
the rate of strain is more important than the largest eigenvalue. 
However, such casual arguments are deceiving~\cite{Tsi2009}, 
since the comparatively rare configurations where vorticity is aligned with the largest eigenvalue of 
the rate of strain may actually contribute more to the dynamics than the frequent
contributions, where vorticity is aligned with the intermediate eigenvector.
Here we check this argument more quantitatively by studying the alignment weighted by the magnitudes of the rate of strain and vorticity.
Again, we study the evolution in the coordinates of $\mathbf{e}_i(0)$, the eigenframe of the rate of strain at some earlier time.

\begin{figure}
\begin{center}
	\includegraphics[width=0.7\textwidth,angle=0]{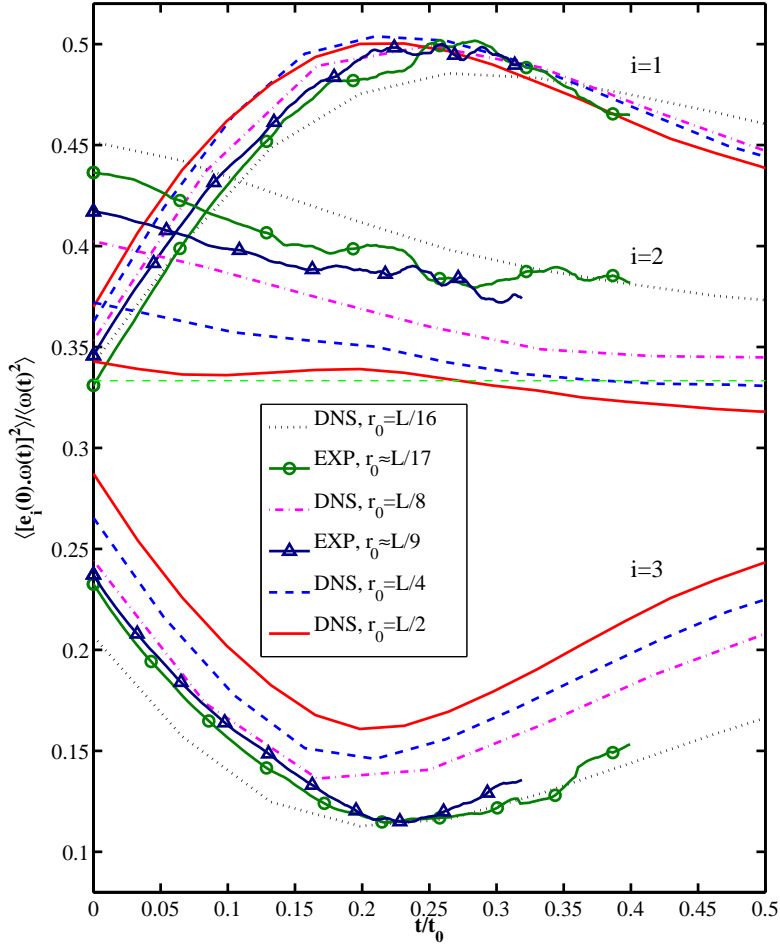}
\caption{(Color online) Alignment of $\mathbf{e}_\omega(t)$ with $\mathbf{e}_i(0)$ weighted by vorticity. 
The averages $\langle [ \mathbf{e}_i(0) \cdot \omega(t) ]^2 \rangle /\langle \omega^2(t) \rangle$
are shown as a function of $t/t_0$ for different values of $r_0$ in the inertial range.
The DNS data correspond to $R_\lambda 
= 170$, and the experiments to $R_\lambda = 350$.
}
\label{fig:align_cond_vort}
\end{center}
\end{figure}

Figure~\ref{fig:align_cond_vort} presents $\langle [ \mathbf{e}_i(0) \cdot \omega(t) ]^2 \rangle /\langle \omega^2(t) \rangle$, {\it i.e.}, the evolution of the vorticity weighted alignment with $\mathbf{e}_i(0)$ (or the relative components of enstrophy in the $\mathbf{e}_i(0)$ direction). 
Compared with Fig.~\ref{fig:dyn_align_eo_ei}a, the contribution in $\mathbf{e}_1(0)$ direction is even larger, a consequence of the increase in vorticity due to vortex-stretching in that direction \cite{XPB11}. The contribution in $\mathbf{e}_2(0)$, on the other hand, decreases with time, which is to be compared with the nearly constant $\langle [ \mathbf{e}_2(0) \cdot \mathbf{e}_\omega(t) ]^2 \rangle$ for $t/t_0 \leq 0.2$ in Fig.~\ref{fig:dyn_align_eo_ei}b.

\begin{figure}
\begin{center}
	\includegraphics[width=0.7\textwidth,angle=0]{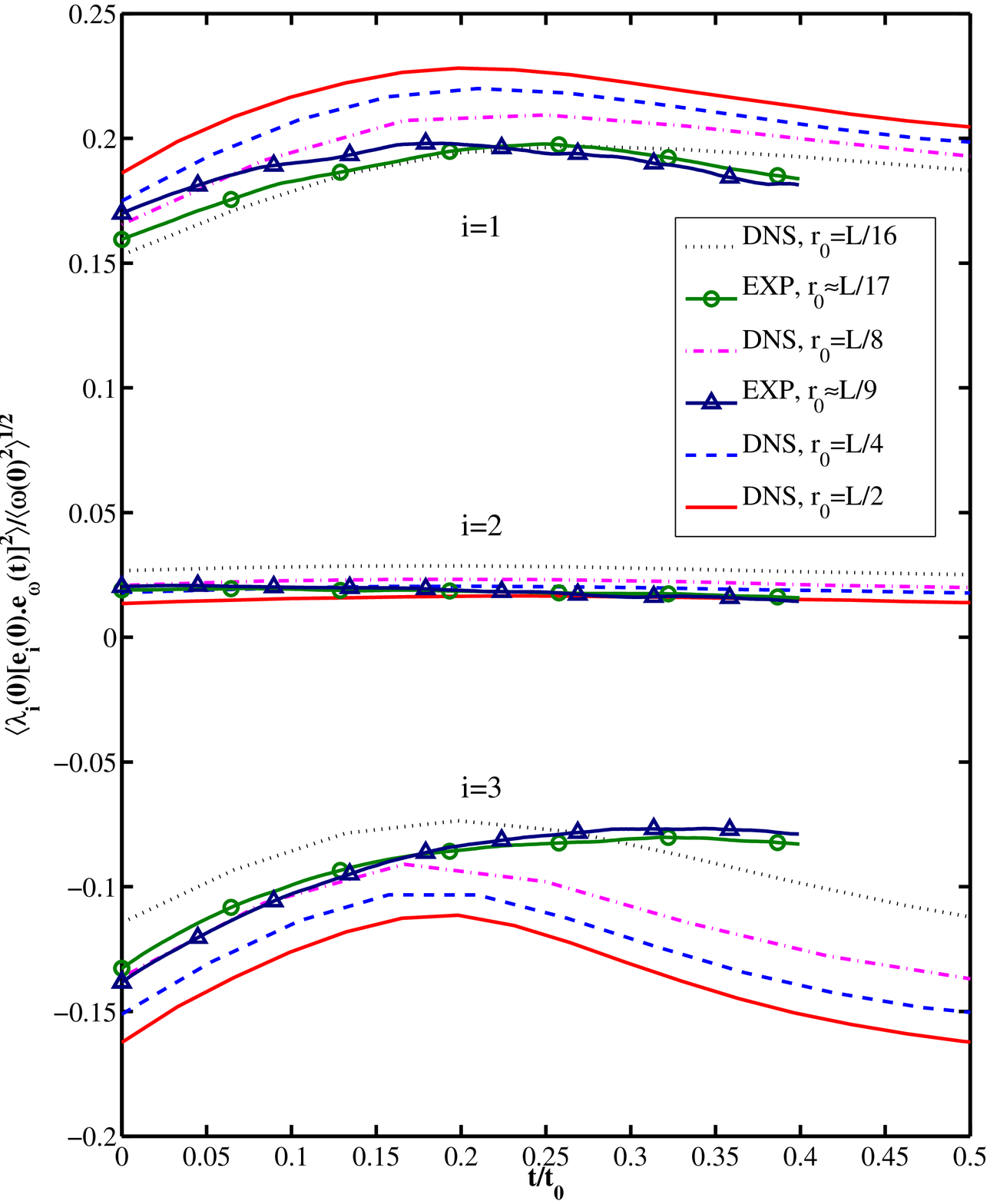}
\caption{(Color online) Alignment of $\mathbf{e}_\omega(t)$ with $\mathbf{e}_i(0)$ weighted by strain. 
The averages $\langle \strain_i(0) [ \mathbf{e}_i(0) \cdot \mathbf{e}_\omega(t) ]^2 \rangle /\langle \omega^2(0) \rangle^{1/2}$
are shown as a function of $t/t_0$ for different values of $r_0$ in the inertial range.
The DNS data correspond to $R_\lambda 
= 170$, and the experiments to $R_\lambda = 350$.
}
\label{fig:align_cond_str}
\end{center}
\end{figure}

The alignment weighted by the magnitude of eigenvalues of the initial rate of strain is shown in Fig.~\ref{fig:align_cond_str}, which calls for the following remarks: 
{\it (i)} The instantaneous alignment is uncorrelated with the magnitude of the rate of strain as $\langle \strain_i(0)[ \mathbf{e}_i(0) \cdot \mathbf{e}_\omega(0) ]^2 \rangle \approx \langle \strain_i(0) \rangle \langle [ \mathbf{e}_i(0) \cdot \mathbf{e}_\omega(0) ]^2 \rangle$. 
{\it (ii)} The increase of $\langle \strain_i(0)[ \mathbf{e}_i(0) \cdot \mathbf{e}_\omega(t) ]^2 \rangle$ with time in the $\mathbf{e}_1(0)$ direction appears to be uncorrelated with the magnitude of $\strain_1(0)$ as the same amount of increase (approximately $30$\%) is achieved in the unconditional statistics $\langle [ \mathbf{e}_1(0) \cdot \mathbf{e}_\omega(t) ]^2 \rangle$ (see Fig.~\ref{fig:dyn_align_eo_ei}a). Similarly, the contribution in $\mathbf{e}_2(0)$ direction is nearly constant at $t/t_0 \leq 0.2$ before decreases, as $\langle [ \mathbf{e}_2(0) \cdot \mathbf{e}_\omega(t) ]^2 \rangle$ in Fig.~\ref{fig:dyn_align_eo_ei}b.

\begin{figure}
\begin{center}
	\includegraphics[width=0.7\textwidth,angle=0]{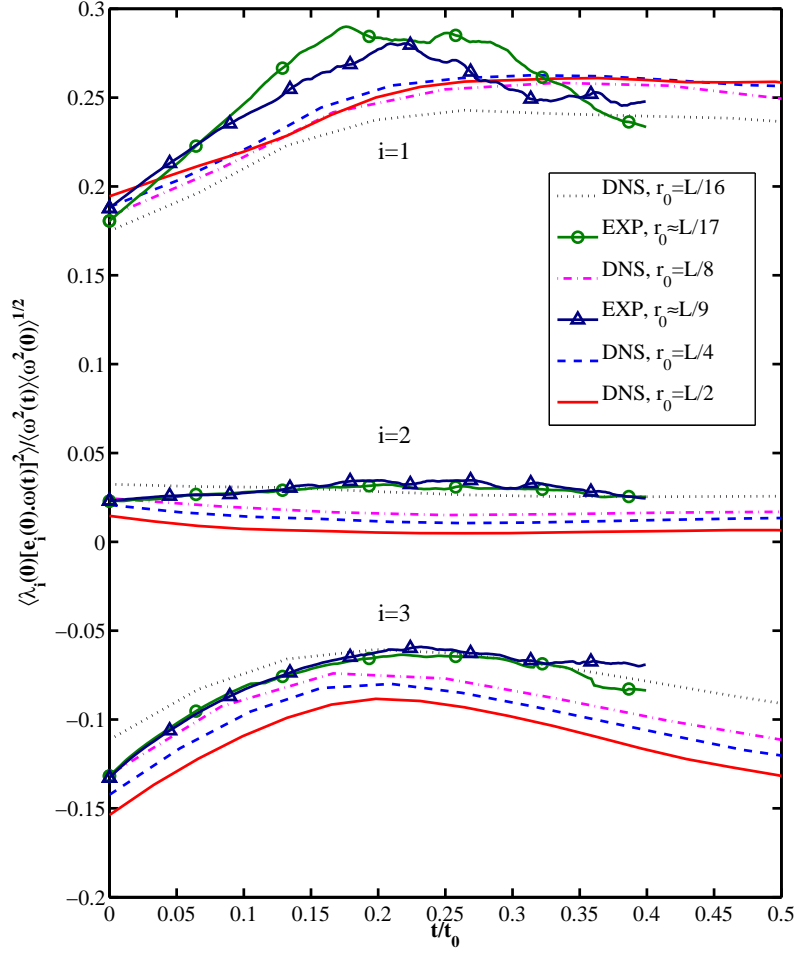}
\caption{(Color online) Alignment of $\mathbf{e}_\omega(t)$ with $\mathbf{e}_i(0)$ weighted by both strain and vorticity. 
The averages
$\langle \strain_i(0) [ \mathbf{e}_i(0) \cdot \omega(t) ]^2 \rangle /\langle \omega^2(t) \rangle \langle \omega(0)^2 \rangle^{1/2}$
are shown as a function of $t/t_0$ for different values of $r_0$ in the inertial range.
The DNS data correspond to $R_\lambda 
= 170$, and the experiments to $R_\lambda = 350$.
}
\label{fig:align_cond_str_vort}
\end{center}
\end{figure}

The quantity 
$\langle \strain_i(0) [ \mathbf{e}_i(0) \cdot \omega(t) ]^2 \rangle$ measures
the evolution of the enstrophy component in the $\mathbf{e}_i(0)$ direction, 
weighted by the eigenvalues of the rate of
strain, $\lambda_i(0)$. At $t = 0$ (instantaneous statistics), the quantity
$\langle \strain_i(0) [ \mathbf{e}_i(0) \cdot \omega(0) ]^2 \rangle$ reduces
to the contribution of the $i^{th}$ eigendirection of $\mathbf{S}$ to 
the vorticity growth: 
$\sum_{i=1}^3 
\langle \strain_i(0) [ \mathbf{e}_i(0) \cdot \omega(0) ]^2 \rangle 
= \langle \omega(0) \cdot \mathbf{S} \cdot \omega(0) \rangle$, which
is the usual vortex stretching.
Fig.~\ref{fig:align_cond_str_vort} thus shows the contribution to vortex-stretching from the directions of $\mathbf{e}_1(0)$ and $\mathbf{e}_2(0)$. 
It is clear that the contribution from $\mathbf{e}_1(0)$ direction increases rapidly with time and is much larger than that from $\mathbf{e}_2(0)$.
 Moreover, the contribution in $\mathbf{e}_1(0)$, once increased, stays at the larger value for longer time as the statistics conditioned on vorticity or the rate of strain alone (see Figs.~\ref{fig:align_cond_vort}a and \ref{fig:align_cond_str}a), which indicates that larger rate of strain $\strain_1(0)$ results in stronger vorticity increase and persists longer.

%
%

%
%

%
%

\subsection{Evolution of the intermediate eigenvalue of the rate of strain}
\label{subsec:evol_interm_strain}

\begin{figure}
\begin{center}
	\includegraphics[width=0.7\textwidth,angle=0]{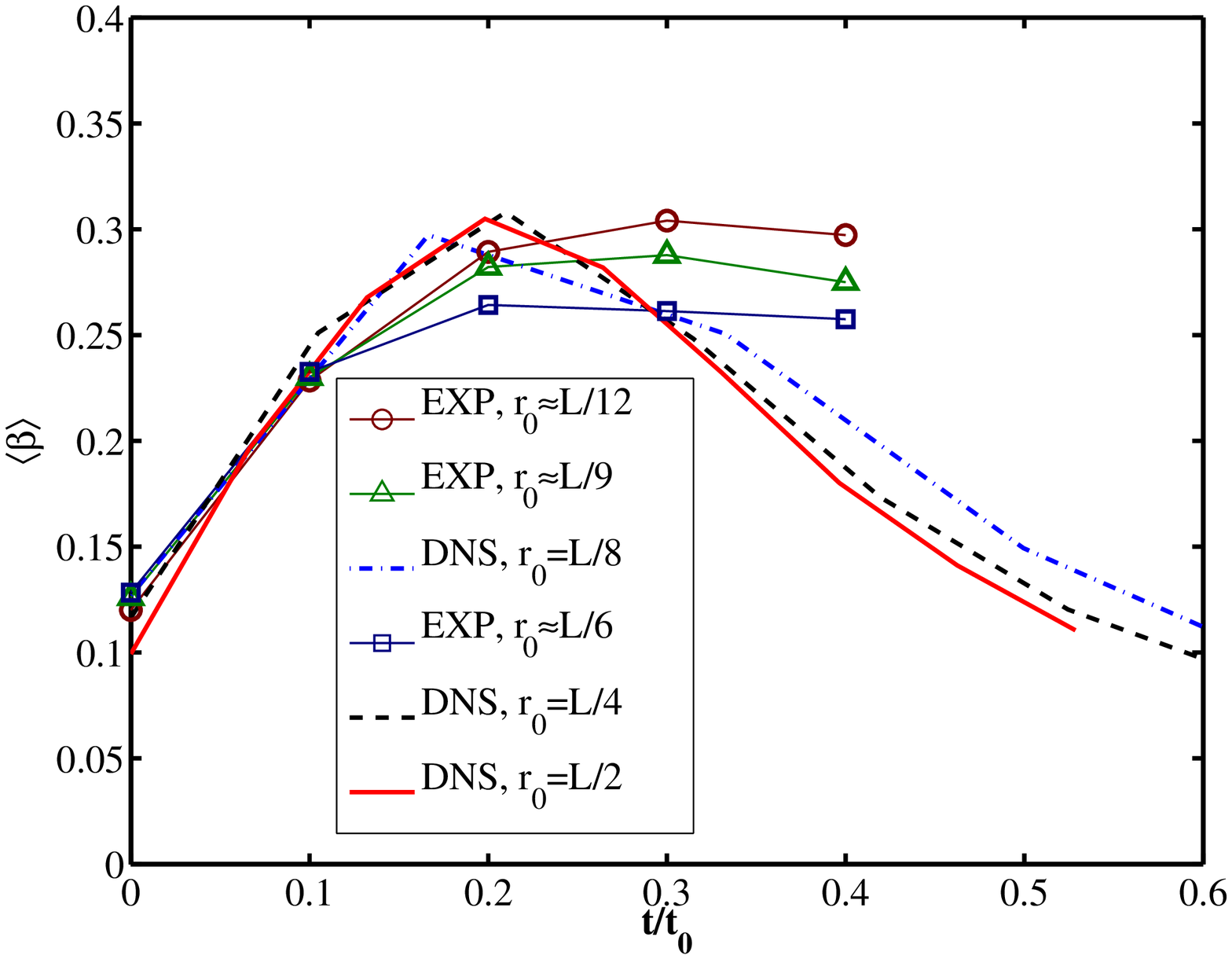}
\caption{(Color online)
Evolution of $\langle \beta \rangle$, the mean of the normalized intermediate eigenvalue of strain. 
With time rescaled with $t_0$, the curves corresponding to different values of $r_0$ collapse very well.  
The DNS data correspond to $R_\lambda 
= 170$, and the experiments to $R_\lambda = 350$.
}
\label{fig:evol_beta}
\end{center}
\end{figure}

In addition to the alignment between vorticity and the eigen-directions of the rate of 
strain, ${\mathbf e}_i$, $\mathbf{M}$ is also characterized by the eigenvalues of the
rate of strain, in particular by the intermediate one, $\lambda_2$. Here, we consider
the normalized value $\beta$, defined by Eq.~\ref{def_beta}.
Fig.~\ref{fig:evol_beta} shows the mean value of $\beta$ as a function
of time $t/t_0$, for several values of the tetrahedron size, $r_0$. 
Remarkably, the curves superpose extremely well for $t/t_0 \lesssim 0.25$, as if the evolution were self-similar.

%
%

\section{Discussion and conclusions } 
\label{sec:discussion}

It is appropriate to stress here that asking questions about
$\mathbf{M}$ is merely one way of studying 
the correlation function of the velocity field $\mathbf{u}(\mathbf{x}, t)$, 
based on four different spatial points.
Although the precise relation between $\mathbf{M}$ and the true
velocity gradient tensor, $\mathbf{m}$, or its coarse-grained generalization is not
completely obvious \cite{Luthi07},
it is our contention that the correlation functions investigated here, based on 
$\mathbf{M}$, contain far more
information than the correlation function of the velocity at two spatial 
points, such as the structure function, and much can be learned about the turbulent flow by investigating correlation
functions at more than two spatial points \cite{Mydlarski98,ShrSig00}.

\subsection{Comparison between dissipative and inertial scale dynamics}
\label{subsec:dissip_inertia}

\begin{figure}
\begin{center}
	\includegraphics[width=0.6\textwidth,angle=0]{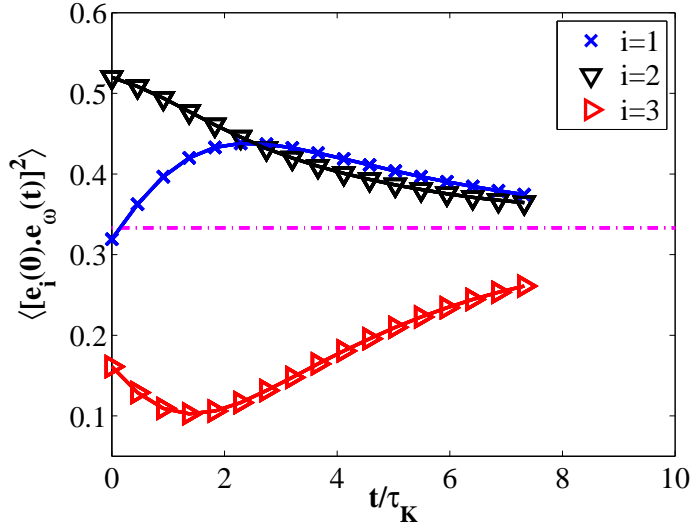}
\caption{(Color online)
Alignment between $\mathbf{e}_\omega(t)$ and $\mathbf{e}_i(0)$ 
for the true velocity gradient tensor $\mathbf{m}$, obtained
from DNS ($R_\lambda = 170$). 
The averages $\langle [\mathbf{e}_i(0) \cdot \mathbf{e}_\omega(t) ]^2 \rangle$
for $i=1$, $2$ and $3$, are indicated by crosses, downwards pointing, and
right pointing triangles. The time is normalized by the Kolmogorov
time scale, $\tau_K = (\nu/\varepsilon)^{1/2}$. The observed evolutions 
of the alignment of ${\bf e}_\omega(t)$ and ${\bf e}_i(0)$ are qualitatively
very similar to what is shown in Fig.~\ref{fig:dyn_align_eo_ei} for tetrahedra
with size $r_0$ in the inertial range. }
\label{fig:dyn_align_eo_ei_dissip}
\end{center}
\end{figure}

In the previous sections, the properties of $\mathbf{M}$ have been studied for 
tetrahedra of size $r_0$ in the inertial range. 
It is appropriate to compare these alignment properties with those obtained 
from the true velocity gradient tensor, $\mathbf{m}$. 
We first note that $\mathbf{M}$ reduces
to the true velocity gradient tensor when $r_0 $ becomes significantly
smaller than the Kolmogorov scale $\eta$.
Here we report results from the DNS on the evolution of the alignment between the
direction of  the true vorticity $\mathbf{e}_\omega(t)$ with the eigenvectors
 $\mathbf{e}_i(0)$ of $\mathbf{s}$. The statistics characterizing the alignment
are shown in Fig.~\ref{fig:dyn_align_eo_ei_dissip}.
The curves in Fig.~\ref{fig:dyn_align_eo_ei_dissip} and 
Fig.~\ref{fig:dyn_align_eo_ei} show strong qualitative similarities.
They differ only in the  scaling of the $t$-axis. 
We find that in the dissipative range the characteristic time scale for alignment between vorticity and strain is $\sim 2-3 \tau_K$, which coincides with the {correlation time } of the rate of strain tensor~\cite{Yeung07,PumWil11}, where $\tau_K \equiv  (\nu/\dissip)^{1/2}$ is the Kolmogorov time scale.
Remarkably, the strongest alignment of ${\bf e}_\omega(t)$ with ${\bf e}_1(0)$ 
in the  dissipative range is as strong as it is in the inertial one: compare
Fig.~\ref{fig:dyn_align_eo_ei_dissip} and Fig.~\ref{fig:dyn_align_eo_ei}(a).
The tendency of ${\bf e}_\omega(0)$ to align with ${\bf e}_2(0)$, and to be perpendicular to ${\bf e}_3(0)$ is stronger
in the dissipative range than in the inertial range, as shown also in Fig.~\ref{fig:pdf_ei_eom}(c).
Thus, the properties of alignment between ${\bf e}_\omega(t) $ and ${\bf e}_i(0)$
for $\mathbf{M}$ when $r_0$ is in the inertial range are quite similar with
the properties of the true velocity gradient
tensor $\mathbf{m}$. Similar observations were reported in another very recent numerical study \cite{ChevMen11}.

%
%

\subsection{Fixed shape tetrahedra} 
\label{subsec:fixed_shape}

The coupling between the dynamics of $\mathbf{M}$ and the geometry of the set
of points has been postulated to be a crucial ingredient in any effort
to properly understand the dynamics of the velocity gradient tensor 
\cite{CPS99,Meneveau11,Chevillard06}. 
The recent experimental and numerical work \cite{XPB11} effectively demonstrates 
that the properties of alignment of vorticity with the eigenvalues of the rate of strain
are strongly dependent on the deformation of the tetrahedra. 
In this context it is worth to ask how the dynamics of alignment is modified,  if instead of following Lagrangian 
tetrahedra that are deformed by the flow, one considers tetrahedra of {\it fixed} shape and size. 
To analyze this we followed numerically one single tracer particle, and used it as the center of mass
of an isotropic tetrahedron with a fixed size and arbitrary orientation.
A similar approach was used before \cite{MenevLund94} for fixed volumes following Lagrangian particles.
For values of $r_0$ much smaller than the Kolmogorov length scale, $\eta$, 
the velocity gradient tensor obtained using this construction reduces to the true velocity gradient 
tensor $\mathbf{m}$, whose properties are illustrated in 
Fig.~\ref{fig:dyn_align_eo_ei_dissip}.

\begin{figure}
\begin{center}
\subfigure[]{
	\includegraphics[width=0.49\textwidth,angle=0]{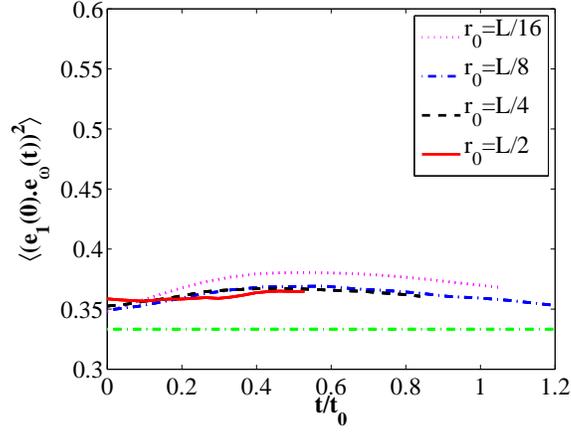}
}
\\
\subfigure[]{
	\includegraphics[width=0.49\textwidth,angle=0]{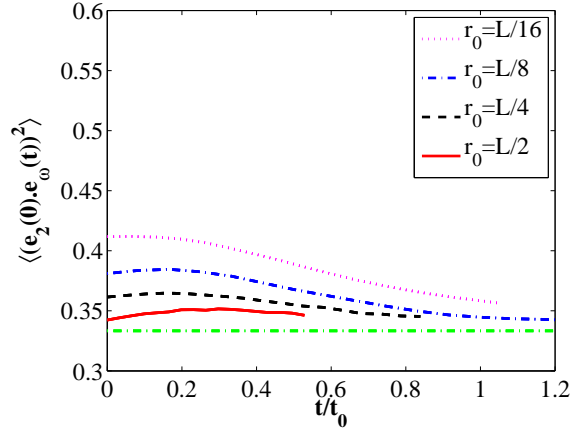}
}
\\
\subfigure[]{
	\includegraphics[width=0.49\textwidth,angle=0]{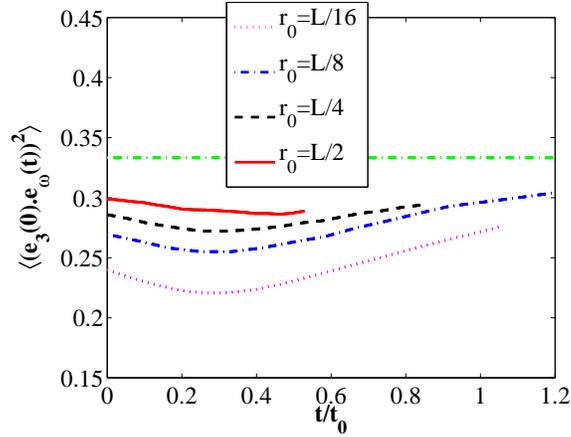}
}
\caption{(Color online)
Alignment between $\mathbf{e}_\omega(t)$ and $\mathbf{e}_i(0)$ 
measured with isotropic tetrahedra of fixed shapes and size $r_0$, obtained from DNS at $R_\lambda = 170$. 
The averages $\langle [\mathbf{e}_i(0) \cdot \mathbf{e}_\omega(t) ]^2 \rangle$
for $i=1$ (a), $2$ (b) and $3$ (c), as a function of $t/t_0$, are shown for several values of 
 $r_0$, all in the inertial range.
The tendency of $\mathbf{e}_\omega(t)$ to align with $\mathbf{e}_1(0)$, and  
to become perpendicular to $\mathbf{e}_3(0)$, are much weaker than with tetrahedra
freely advected with the flow (compare with Fig.~\ref{fig:dyn_align_eo_ei}), 
or with the true velocity gradient tensor (compare with Fig.~\ref{fig:dyn_align_eo_ei_dissip}).
}
\label{fig:dyn_align_eo_ei_fix}
\end{center}
\end{figure}

Fig.~\ref{fig:dyn_align_eo_ei_fix} shows the evolution of the alignment of 
${\mathbf e}_\omega(t)$ with the eigenvectors of the rate of strain ${\mathbf e}_i(0)$
for the velocity gradient tensor obtained from  {\it fixed shape } tetrahedra.
Comparing Fig.~\ref{fig:dyn_align_eo_ei} and Fig.~\ref{fig:dyn_align_eo_ei_fix}
reveals some strong quantitative differences. The tendency 
${\bf e}_\omega(t)$ to align with ${\bf e}_1(0)$, as measured by the increase 
of $\langle ( {\bf e}_\omega(t) \cdot {\bf e}_1(0) )^2 \rangle$, is much weaker for fixed-shape tetrahedra than for deformable 
tetrahedra. Similarly, the tendency of ${\bf e}_\omega(t)$ to become more
perpendicular to ${\bf e}_3(0)$ is much reduced. 
We also notice that the time scale {of the alignment process} is much longer for fixed{{-shape}} tetrahedra. 
Moreover, the {degrees of} alignment for fixed-shape tetrahedra (Fig.~\ref{fig:dyn_align_eo_ei_fix}) 
are much weaker than for the dissipative range.

The results shown in Fig.~\ref{fig:dyn_align_eo_ei_fix} thus demonstrate a very important feature:
the properties of alignment of $\mathbf{M}$, constructed with 
properly deformable tetrahedra are much closer to those of the true
velocity gradient tensor, than those obtained with fixed shape
tetrahedra. The coupling between
the dynamics of $\mathbf{M}$ and of the geometry is crucial to properly 
understand and model fluid motion.

%
%

\subsection{Summary}
\label{subsec:sum_concl}

To sum up, we have investigated systematically the perceived velocity gradient $\mathbf{M}$ from the point
of view of alignment between vorticity and the rate of strain, using both
DNS at $R_\lambda = 170$, over the entire range
of scales, and experiments at a higher Reynolds numbers 
($R_\lambda = 350$) but 
over a more restricted range of scales in the inertial range. Our results 
generalize our recent work \cite{XPB11}, which focused on the
alignment between vorticity and the eigenvector ${\mathbf e}_1$ corresponding
to the largest stretching. 

We found that strong deformation of tetrahedra, leading to the formation
of flattened, almost coplanar configurations,
occurs over a time scale of the order $t_0/4$, {which is comparable to the
time scale }
characterizing the alignment of vorticity with the {eigenvector
$\mathbf{e}_1(0)$ of the strain, corresponding to the largest eigenvalue, 
$\sim t_0/5$, see Fig.~\ref{fig:dyn_align_eo_ei}(a)}. 
{We note in this respect that during this time, the relative
orientation of the vectors 
$\mathbf{e}_\omega(t)$ and $\mathbf{e}_2(0)$ remains nearly constant, see Figs.~\ref{fig:dyn_align_eo_ei}(b) and \ref{fig:angle_ei_eom}(b), (e), and (h)}.

Whereas the dynamics leading to alignment between 
$\mathbf{\omega}$ and ${\mathbf e}_1$ has been found to be essentially 
self-similar \cite{XPB11} over the whole range of scales {\it and} Reynolds numbers considered, 
the self-similarity of the statistical properties of the angle between
$\omega$ and the other two eigendirections of the rate of strain
is only observed at the largest Reynolds number studied here, $R_\lambda = 350$.
At small Reynolds numbers, $R_\lambda \le 170$, we found a systematic 
dependence on the size $r_0$ of the tetrahedra of the alignment of vorticity and the rate of strain at the \textit{same} time. 
We also observed that the evolution
of the angle between vorticity and the second and third eigenvalues of the rate of strain
does not simply reduce to a simple function of $t/t_0$.
In any event, the dynamics proceed with a characteristic time scale of order 
{$\sim t_0/5$}.

For the range of Reynolds numbers considered here, the picture
that emerges from the study of $\mathbf{M}$ reflects the properties in the 
dissipative scale in terms of alignment of vorticity and strain eigenvectors.
The characteristic time of evolution in the dissipative scale is $\sim 2-3 \tau_K$. 

Future studies will be devoted to a more systematic comparison with modeling,
both in terms of the tetrahedron model\cite{CPS99}, and in terms of the
arguments developed in our recent work \cite{XPB11}.

\section*{acknowledgement}
We are very thankful to B. L\"uthi, B. Shraiman, E. Siggia and A. Tsinober for stimulating 
discussions, and to L. Chevillard and C. Meneveau 
for communicating to us reference \cite{ChevMen11}, and for freely sharing 
their insight.
AP thanks IDRIS for providing the computation resources and the ANR for financial support through the contract TEC2. 
HX is grateful to the Deutsche Forschungsgemeinschaft for support through the grant XU 91/3-1.
We thank the Max Planck Society for support. This research was conducted in part at the Kavli Institute for Theoretical Physics at Santa Barbara, CA (supported by the US National Science Foundation under Grant No. NSF PHY05-51164), the Kavli Institute for Theoretical Physics China at Beijing (supported by the Project of Knowledge Innovation Program (PKIP) of the Chinese Academy of Sciences under Grant No.~KJCX2.YW.W10) and received additional support from the the European COST Action MP0806.

\end{document}